\documentclass[letterpaper,english,aps,prb,floatfix,showpacs,amsfonts,amssymb,superscriptaddress]{revtex4}
\usepackage[T1]{fontenc}
\usepackage[latin1]{inputenc}
\usepackage{graphics}
\usepackage{amssymb}
\usepackage{amsmath}
\usepackage{overpic}

\newcommand{\be}{\begin{equation}}
\newcommand{\ee}{\end{equation}}
\newcommand{\bea}{\begin{eqnarray}}
\newcommand{\eea}{\end{eqnarray}}

\newcommand{\rmd}{{\rm d}}
\newcommand{\rmi}{{\rm i}}

\newcommand{\cuoo}{CuO$_{2}$}
\newcommand{\lco}{La$_{2}$CuO$_{4}$}
\newcommand{\lasco}{La$_{2-x}$Sr$_{x}$CuO$_{4}$}
\newcommand{\srcuocl}{Sr$_2$CuO$_2$Cl$_2$}

\renewcommand\a{\alpha}
\renewcommand\b{\beta}

\renewcommand\d{\delta}

\renewcommand\t{\tau}

\newcommand\s{\sigma}

\newcommand\ra{\rightarrow}

\newcommand\pll{\parallel}

\newcommand\Ra{\Rightarrow}
\newcommand\pd{\partial}
\newcommand\nb{\nabla}
\newcommand\bk{{\bf k}}
\newcommand\bq{{\bf q}}
\newcommand\bQ{{\bf Q}}

\newcommand\bH{{\bf H}}
\newcommand\bD{{\bf D}}
\newcommand\bL{{\bf L}}

\newcommand\bS{{\bf S}}
\newcommand\bn{{\bf n}}
\newcommand\bx{{\bf x}}

\newcommand\LL{{\cal L}}

\newcommand\cA{{\cal A}}

\newcommand\nn{\nonumber}
\newcommand\lb{\label}
\def\pref#1{(\ref{#1})}

\newcount\bozza \bozza=0
\ifnum\bozza=1
\newdimen\shift \shift=-2truecm
\def\lb#1{%
{\label{#1}\rlap{\kern\shift{$\scriptstyle#1$}}}}
\else\def\lb#1{\label{#1}} \fi

\begin{document}

\title{Field dependence of the magnetic spectrum in
  anisotropic and Dzyaloshinskii-Moriya antiferromagnets: I. Theory}

\author{L.~Benfatto}

\email{lara.benfatto@roma1.infn.it}

\affiliation
{CNR-SMC-INFM and Department of Physics, University of Rome ``La
  Sapienza'',\\ Piazzale Aldo Moro 5, 00185, Rome, Italy}

\author{M.~B.~Silva~Neto}

\email{barbosa@phys.uu.nl}

\affiliation {Institute for Theoretical Physics, University of Utrecht,
  P.O. Box 80.195, 3508 TD, Utrecht, The Netherlands}

\date{\today}

\begin{abstract}

We consider theoretically the effects of an applied uniform magnetic field
on the magnetic spectrum of anisotropic two-dimensional and
Dzyaloshinskii-Moriya layered quantum Heisenberg antiferromagnets. The
former case is relevant for systems such as the two-dimensional square
lattice antiferromagnet {\srcuocl}, while the latter is known to be
relevant to the physics of the layered orthorhombic antiferromagnet
{\lco}. We first establish the correspondence between the low-energy
spectrum obtained within the anisotropic non-linear sigma model and by
means of the spin-wave approximation for a standard easy-axis
antiferromagent. Then, we focus on the field-theory approach to calculate
the magnetic-field dependence of the magnon gaps and spectral intensities
for magnetic fields applied along the three possible crystallographic
directions. We discuss the various possible ground states and their
evolution with temperature for the different field orientations, and the
occurrence of spin-flop transitions for fields perpendicular to the layers
(transverse fields) as well as for fields along the easy axis (longitudinal
fields). Measurements of the one-magnon Raman spectrum in {\srcuocl} and
{\lco} and a comparison between the experimental results and the
predictions of the present theory will be reported in part II of this
research work [L.~Benfatto {\it et al.}, forthcoming article, 
cond-mat/0602664].

\end{abstract}


\maketitle

\section{Introduction}

The relevance of antisymmetric superexchange interactions in
spin Hamiltonians describing quantum antiferromagnetic (AF)
systems has been acknowledged long ago by Dzyaloshinskii.\cite{D}
Soon after, Moriya showed that such interactions arise naturally
in perturbation theory due to the spin-orbit coupling in magnetic
systems with low symmetry.\cite{M} Nowadays, a number of AF
systems are known to belong to the class of Dzyaloshinskii-Moriya
(DM) magnets, such as $\alpha$-Fe$_2$O$_3$, LaMnO$_3$,\cite{Talbayev} and
K$_2$V$_3$O$_8$,\cite{Lumsden} and to exhibit unusual and interesting magnetic
properties in the presence of quantum fluctuations and/or applied
magnetic field.\cite{Talbayev,Kotov,sasha}

Also belonging to the class of DM antiferromagnets is {\lco}, which is a
parent compound of high-temperature superconductors. In {\lco} the unique
combination of antisymmetric superexchange, caused by the staggered tilting
pattern of oxygen octahedra around each copper ion in the low-temperature
orthorhombic (LTO) phase, and weak interlayer coupling, results in an
interesting four sublattice structure for the antiferromagnetism of {\lco},
where the Cu$^{++}$ spins are canted out of the CuO$_2$ layers with
opposite canting directions between neighboring
layers.\cite{Shekhtman,Thio,Thio90,Papanicolaou} These small ferromagnetic
moments lead to a quite unconventional physics in the antiferromagentic
phase. For example, when a small magnetic field is applied perpendicular to
the layers the magnetic susceptibility shows a strong enhancement as the
Ne\'el temperature is approached, due to the formation of these
ferromagnetic moments.\cite{Thio,MLVC,Gooding}. When the field strenght is
further enhanced one can observe a spin flop of the ferromagnetic moments
whith respect to the out-of-plane staggered order that they display at zero
field. \cite{Thio,Ando-Mag-Anisotropy,Papanicolaou} Analogously the
spin-flop transition of the staggered in-plane AF moments for a field along
the easy axis is accompanied by new effects related to the presence of the
DM interaction. \cite{Thio90,Papanicolaou} Recently new theoretical
approaches have been proposed \cite{MLVC,Gooding,Raman} to integrate the
semiclassical picture already presented in the prior work of
Refs. [\onlinecite{Shekhtman,Thio,Thio90,Papanicolaou}]. Even thought the basic
physical picture remains the same, the inclusion of quantum effects in the 
long-wavelenght formulation of the spin problem discussed in
Ref. [\onlinecite{MLVC,Raman}] allowed for a straightforward and complete
understanding of the unusual magnetic-susceptibility anisotropies observed
in {\lco} for a rather large temperature range, $0<T<400$ K. Moreover, a
particular attention has been devoted in Ref.\ [\onlinecite{Raman}] to the
analysis of the one-magnon excitations by means of Raman spectroscopy,
and the use of the long-wavelength theory turns out to be very convenient
to understand why the DM interaction is behind the appearance\cite{Gozar}
of a field-induced mode for an in-plane magnetic field.\cite{Raman}

A better understanding of the anomalies related to the presence of the DM
interaction in {\lco} compounds can be achieved by directly comparing its
properties with those of a similar spin system like {\srcuocl}.  In this
case the DM interaction is absent due to the higher crystal symmetry, but
spin-orbit coupling can still give rise to small anisotropies of purely
quantum mechanical origin.\cite{yildrim} Thus quantum corrections coming
from spin-orbit coupling give rise to a quite small easy-axis anisotropy,
so that a gap in the in-plane magnon excitations has been observed in
ESR.\cite{ESR} As a consequence, {\srcuocl} behaves as an ordinary
easy-axis antiferromagnet, in contrast to {\lco} which should be classified
as an unconventional easy-axis antiferromagnet. 

It is the purpose of this article to study in detail the influence of an
applied uniform magnetic field on the magnetic spectrum for the above two
cases: anisotropic two-dimensional and layered DM antiferromagnets, and to
compare the qualitative differences between these two cases. This will be
done mainly using the continuum quantum field theory appropriate for each
of these two cases, i.e. the non-linear sigma model (NLSM), properly
modified to account for conventional or DM anisotropies. Nonetheless, we
will sketch in the beginning the calculation of the magnon gaps within the
framework of the semiclassical approximation for conventional
antiferromagnets, and we demonstrate the complete equivalence between the
two approaches as far as the gaps values at low temperature are
concerned. However, as it will become clear in the following, the quantum
NLSM followed here allows also to account for the quantum and thermal
effects of the spin fluctuation, which were neglected in the previous
approaches \cite{Thio,Thio90,Papanicolaou}. As a consequence, we can
evaluate (within the given saddle-point approximation for the transverse
spin fluctuations) the full $(H,T)$ phase diagram for a field in the
various direction, which can be compared with the existing experimental
data.  At the same time, the continuum field theory provides an elegant and
straighforward description of the spin fluctuations in the coupled-layers
case, and also of the various spin-flop transitions that may occur in
{\lco} at moderate fields.

The structure of the paper is as follows. In Section II we introduce the
model Hamiltonians for {\srcuocl} and {\lco}, appropriate to describe a
conventional anisotropic two-dimensional antiferromagnet and a DM
antiferromagnet, respectively. In Section III we sketch the standard
semiclassical calculations of the magnon gaps at zero field in the two
cases, and at finite magnetic field for the standard anisotropic case. Then
the same (conventional) results are reproduced in Section IV using the NLSM
approach, whose main properties are here described. Section V is dedicated
to the case of a layered DM antiferromagnet, and the effects of an uniform
magnetic field applied along the three crystallographic directions are
extensively discussed, with reference to the specific structure of {\lco}.
The conclusions are reported in Section VI.
In a second part,\cite{theo-exp} we shall make a quantitative
comparison between the predictions of the theory developed in this
article and the magnetic spectrum probed by one-magnon Raman
scattering in both {\lco} and {\srcuocl}.

\section{Spin-Hamiltonians for conventional and DM anisotropies}

\begin{figure}[htb]
\includegraphics[scale=0.4]{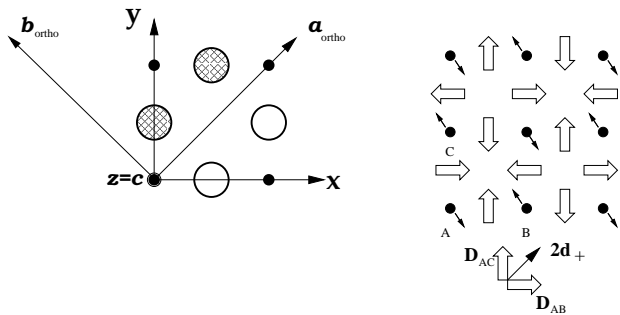}
\caption{Left: the hatched circles represent the O$^{--}$ ions tilted above
  the {\cuoo} plane; the empty ones are tilted below it; small black
  circles are Cu$^{++}$ ions. Right: Schematic arrangement of the staggered
  magnetization (small arrows) and DM vectors (open arrows). Right
  bottom: definition of the vector ${\bf d}_{+}= {\bf D}_+/4S$.}
\label{Fig-1}
\end{figure}

{\lco} is a body centered orthorhombic antiferromagnet with $Bmab$
crystal structure. A single layer of {\cuoo} ions in {\lco} can be
described by the $S=1/2$ Hamiltonian
\be
\lb{Hamiltonian}
H_{sl}[{\bf S},{\bf D}]=J\sum_{\langle i,j\rangle}{\bf S}_{i}\cdot{\bf S}_{j}+
\sum_{\langle i,j\rangle}{\bf D}_{ij}\cdot\left({\bf S}_{i}\times{\bf
    S}_{j}\right)+
\sum_{\langle i,j\rangle}{\bf S}_{i}
\cdot\overleftrightarrow{\bf \Gamma}_{ij}\cdot{\bf S}_{j},
\ee
where ${\bf D}_{ij}$ and $\overleftrightarrow{\bf \Gamma}_{ij}$
are, respectively, the DM and XY anisotropic interaction terms
that arise due to the spin-orbit coupling and direct-exchange in
the low-temperature orthorhombic (LTO) phase of
{\lco}.\cite{Shekhtman} Throughout this work we adopt the LTO
$(abc)$ coordinate system of Fig.\ \ref{Fig-1},\cite{coord-system}
for both the spin and lattice degrees of freedom, and we use units
where $\hbar=k_{B}=1$.

The direction and the alternating pattern of the DM vectors, shown in Fig.\
\ref{Fig-1}, have been calculated by several authors \cite{Shekhtman} by
taking into account the tilting structure of the oxygen octahedra and of
the symmetry of the {\lco} crystal. For {\lco} the DM vectors are in good
approximation perpendicular to the $Cu-Cu$ bonds and change sign from one
bond to the next one:
\be \lb{dmvectors} {\bf D}_{AB}=\frac{1}{\sqrt{2}}(d,-d,0), \quad
{\bf D}_{AC}=\frac{1}{\sqrt{2}}(d,d,0), \ee
while the $XY$ matrices $\overleftrightarrow \Gamma$ provide essentially an
easy-plane anisotropy for the Hamiltonian \pref{Hamiltonian}:
\bea
\lb{eq:Gamma}
\overleftrightarrow{\Gamma}_{AB}{=}
  \left( \begin{array}{ccc}
     \Gamma_1 & \Gamma_2 & 0 \\
     \Gamma_2 & \Gamma_1 & 0 \\
     0 & 0 & \Gamma_3
         \end{array} \right)\!\!,\;
\overleftrightarrow{\Gamma}_{AC}{=}
  \left( \begin{array}{ccc}
     \Gamma_1 & -\Gamma_2 & 0 \\
     -\Gamma_2 & \Gamma_1 & 0 \\
     0 & 0 & \Gamma_3
         \end{array} \right)\!\!,
\nonumber
\eea
where $AB$ and $AC$ label the Cu$^{++}$ sites on horizontal/vertical bonds
respectively (see Fig.\ \ref{Fig-1}). As it has been stressed by Shekhtman
et al. \cite{Shekhtman}, even though the parameters $d$ and
$\Gamma_{1,2,3}>0$ have different orders of magnitude, with $d\sim
10^{-2}J$ and $\Gamma_i\sim 10^{-4}J$, they should be considered on the
same footing (see also discussion following Eq.\ \pref{gap-zf}
below). Indeed, one can show that considering the two last terms of Eq.\
\pref{Hamiltonian} the interaction between spins on a neighboring bond can
be written in a completely isotropic form by rotating locally the spin
operators around the ${\bf D}_{ij}$ axis by an angle $\theta_{ij}=\arctan
|{\bf D}_{ij}|/2J$. As a consequence, weak ferromagnetism arises only when
global frustration of the DM pattern exists. In term of the DM vectors
defined above, this condition corresponds to having ${\bf d}_+\neq {\bf
d}_-$, where ${\bf d}\pm=({\bf D}_{AB}\pm{\bf D}_{BC})/2$. This condition
is clearly satisfied by the DM vectors \pref{dmvectors}, so that WF is
expected in {\lco}.

A realistic model for {\lco} should include also interlayer coupling. In
the orthorhombic unit cell of {\lco} the spins of the $Cu$ atoms are
displaced by an in-plane diagonal vector $(1/2,1/2,0)$ from one layer to
the next one. As a consequence, given a couple of spins in a layer and the
nearest couple in the next one, the DM vector of the corresponding bond
will change sign. We can then write the full Hamiltonian as:
\be \lb{htot} H=J_\perp\sum_m {\bf S}^m\cdot{\bf S}^{m+1}+\sum_m
H_{sl}[{\bf S}^m,{\bf D}^m], \ee
where ${\bf S}^m$ represents the spin at a generic position $(i,j)$ of the
$m$th plane and ${\bf D}^m_{AB,AC}=(-1)^m{\bf D}_{AB,AC}$. Since the {\lco}
unit cell is body centered, the coupling $J_\perp$ in Eq.\ \pref{htot}
connects the two spins at $(0,0,0)$ and $(1/2,0,1/2)$ in the LTO
notation. It is worth noting that the pure 2D system \pref{Hamiltonian}
does not display any spin rotational symmetry, so it can order at finite
temperature without violating the Mermin-Wagner theorem. However, in the
presence of an interlayer coupling the transition to the AF state will
ultimately have 3D character, with spins aligned AF also in the $c$
direction. On the other hand, as we shall discuss below, the interplay
between the interlayer coupling and the DM interaction can lead to a quite
unconventional behavior in the presence of a finite magnetic field. Indeed,
the DM interaction is not only the source of an easy-plane anisotropy (the
spins prefers to align perpendicularly to the DM vector ${\bf d}_+$), but
induces also an anomalous coupling between the AF order parameter and an
applied magnetic field.

These effects can be better understood by comparing the results
obtained with the model \pref{Hamiltonian} (and its
three-dimensional version \pref{htot}) with the ones coming from a
more conventional 2D anisotropic Heisenberg model, as:
\be
\lb{conv}
H_{con}=\sum_{\langle i,j\rangle}JS_i^bS_j^b+(J-\alpha_a)S_i^aS_j^a+
(J-\alpha_c)S_i^cS_j^c.
\ee
The Hamiltonian \pref{conv} is the appropriate starting model for
{\srcuocl}, where interlayer coupling is even less relevant than in {\lco}
due to the frustation on the tetragonal unit cell. Here the
crystallographic in-plane $a,b$ axes are choosen with $b$ parallel to the
spin easy axis (which is along the $xy$ direction), so that we will have a
similar notation to the one used for {\lco}. However, $a=b$ for {\srcuocl},
since the system is tetragonal. As we explained in the introduction,
$\a_a$ should be zero in a tetragonal system. Nonetheless, quantum effects
can induce an in-plane anisotropy\cite{yildrim} which we will mimic with a
finite $\a_a$ anisotropy term in what follows. To clarify to what extent
the DM interaction introduces an anomalous behavior, we shall start our
analysis of easy-axis antiferromagnetism from the anisotropic model
\pref{conv}. We will then be able to go back to the model
\pref{Hamiltonian}-\pref{htot} and to correctly distinguish the effects of
the magnetic field alone from the ones arising from the presence of the DM
interaction.

%
%
%

\section{Semiclassical approach}

\subsection{Conventional anisotropies}

Let us start our analysis of the single-layer Hamiltonian
\pref{conv} using a semiclassical approach. Similar calculations
have been already carried out in different contexts,
\cite{kittel,Kubo} and here we shall just review the main steps to
fix the notation and to clarify the conventional behavior of an
ordinary easy-axis antiferromagnet.  Let us denote by $\bS_1$ and
$\bS_2$ the spins on the two AF sublattices.  The free energy density in
the AF phase can be written:
\be
\lb{anis}
F=zJS_1^bS_2^b+z(J-\a_a)S_1^aS_2^a+z(J-\a_c)S_1^cS_2^c,
\ee
where $z$ is the number of nearest neighbors and $S^{a,b,c}$ are the
components of the vector along the three crystallographic axes.  Within the
semiclassical approach, the spins are treated as classical vectors of
length $S$: thus the ground-state configuration can be easily determined by
imposing that $\nabla_{\bS_i}F=0$. This condition clearly shows that the
spins order along the $b$ direction, with $\bS_1^0=-\bS_2^0=S\bx_b$.  To
calculate the magnon gaps one uses the classical equations of motion:
\be
\lb{equations}
\frac{d\bS_i}{dt}=\bS_i\times \nabla_{\bS_i}F, \quad i=1,2
\ee
where $\nabla_{\bS_i}F/(g_s\mu_B)$ represents the effective local
magnetic field around which each magnetic moment $(g_s\mu_B)\bS$
precesses. Here $g_s$ is the gyromagnetic ratio and $\mu_B$ is the
Bohr magneton. By expanding the spins around the classical solution,
$\bS_1=S(m_1^a,1,m_1^b),
\bS_2=S(m_2^a,-1,m_2^b)$ Eqs.\ \pref{equations} give a set of 4 coupled
equations for the time dependence of the transverse spin fluctuations,
which can be easily solved by putting $m_i^\a(t)= m_i^\a \exp(i\omega
t)$. It then follows that the $\omega$ are the eigenvalues of the matrix
(corresponding to the vector $(m_1^c,m_2^c,m_1^a,m_2^a)$):
\be
\lb{nofield}
i zS \left( \begin{array}{cccc}
0 & 0 & J & J-\a_c \\
0 & 0 & -J+\a_c & -J \\
-J & -J+\a_a & 0 & 0 \\
J-\a_a & J & 0 & 0
         \end{array} \right).
\ee
The eigenvalues $\omega=\pm \omega_a$ and $\omega=\pm \omega_c$ correspond
to eigenmodes describing spin fluctuations with a larger $a/c$ component
respectively, allowing for the identification of $\omega_a$ and $\omega_c$
as the magnon gaps for the in-plane and out-of-plane spin-wave
modes:\cite{kittel}
\be
\lb{gap-zf-anis}
\omega_a=m_a=zS\sqrt{2\a_aJ},
\quad \omega_c=m_c=zS\sqrt{2\a_cJ},
\ee
where we approximated $\sqrt{2\alpha_{a,c}J-\alpha_a\alpha_c}\approx
\sqrt{2\alpha_{a,c}J}$ because $\alpha_a,\alpha_c\ll J$.

In the presence of a finite magnetic field a term $-\sum_i \bH \cdot
\bS_i$ must be added to Eq.\ \pref{conv}, which translates into a term
$-\bH \cdot (\bS_1+\bS_2)$ in Eq.\ \pref{anis}. Observe that in what
follows we shall measure the magnetic field in units of $g_s\mu_B=1$,
unless explicitly stated.  Then one follows the same procedure as
before, by noticing that when a {\em transverse} field is applied (i.e. a
field perpendicular to the easy axis) the sublattice ground-state
configurations $\bS^0_{1,2}$ acquire an uniform component in the field
direction proportional to $H/z(2J-\a_{a,c})$ (for a field along $a$
and $c$, respectively). By adding fluctuations transverse to the new
equilibrium direction one finds for example for $\bH \pll a$ the
fluctuation matrix:
$$
i zS \left( \begin{array}{cccc}
0 & 0 & J & J-\a_c \\
0 & 0 & -J+\a_c & -J \\
-J\cos^2\phi+(J-\a)\sin^2\phi-(H/zS)\sin\phi & J\sin^2\phi-(J-\a_a)\cos^2\phi
& 0 & 0 \\
-J\sin^2\phi+(J-\a)\cos^2\phi  & J\cos^2\phi-(J-\a)\sin^2\phi+(H/zS)\sin\phi
& 0 & 0
         \end{array} \right),
$$
where $\phi=\arcsin(H/zS(2J-\a_a))\approx H/zS2J$. As a consequence,
the new magnon gaps are:\cite{Kubo}
\be
\lb{hplla}
\omega_a(H)=\sqrt{m^2_a+H^2}, \quad \omega_c(H)=m_c.
\ee
Observe that since the canting of the spins due to the magnetic field is
small, the eigenmodes still describe fluctuations having predominantly $a$
or $c$ character, respectively. We see that the effect of a transverse
magnetic field is to harden the gap of the mode in the field direction, and
to leave the other gap unchanged. Indeed, when $\bH$ is parallel to $c$ we
find a similar result, with an increasing $\omega_c$ gap and a constant $\omega_a$
gap.

Finally, let us consider the case of a {\em longitudinal} field, i.e. of a
field parallel to the easy axis. In this configuration no uniform spin
magnetization develops, but the magnetic field effectively shifts the AF
coupling along the easy axis in the two sublattices, so that the new
fluctuation matrix reads:
$$
i zS \left( \begin{array}{cccc}
0 & 0 & J+H/zS & J-\a_c \\
0 & 0 & -J+\a_c & -(J-H/zS) \\
-(J+H/zS) & -J+\a_a & 0 & 0 \\
J-\a_a & J-H/zS & 0 & 0
         \end{array} \right).
$$
The four eigenvalues are given by
$$
{\omega^2}/{(zS)^2}=J(\alpha_a+\alpha_c)-\alpha_a\alpha_c+(H/zS)^2
\pm\sqrt{(\alpha_c-\alpha_a)^2J^2+
4(\alpha_a+\alpha_c)J(H/zS)^2-(H/zS)^2(\alpha_a+\alpha_c)^2},
$$
and using the fact that $\alpha_a,\alpha_c\ll J$ they can be readily expressed
in terms of the bare gaps $m_a,m_c$ as:
\bea
{\omega_a^2}=\frac{m^2_a+m_c^2}{2}+H^2-
\sqrt{\left(\frac{m^2_a-m_c^2}{2}\right)^2+4H^2
  \left(\frac{m^2_c+m_a^2}{2}\right)}, \nn\\
\lb{hpllb}
{\omega_c^2}=\frac{m^2_a+m_c^2}{2}+H^2+
\sqrt{\left(\frac{m^2_a-m_c^2}{2}\right)^2+4H^2
  \left(\frac{m^2_c+m_a^2}{2}\right)}, 
\eea
where we assumed $m_c>m_a$. Note that at fields larger than the bare
gaps one observes essentially a linear increase of the magnon gaps with
the magnetic field. In the case of
degenerate gaps, $\a_c =\a_a$, only the linear regime is accessible and
Eqs. \pref{hpllb} simplify to:\cite{kittel}
$$
\omega_a=m_a-H, \quad \omega_c=m_c+H.
$$

\subsection{Dzyaloshinskii-Moriya interactions}

Let us discuss how the previous results are modified in the presence of DM
interactions. First, taking into account that $\Gamma_{1,2}\ll d\ll J$ the
free energy density corresponding to the Hamiltonian \pref{Hamiltonian} can
be written as:
$$
F=zJ(S_1^aS_2^a+S_1^bS_2^b)+z(J-\a_c)S_1^cS_2^c+zd_+(S_1^bS_2^c-S_1^cS_2^b),
$$
where $\a_c=\Gamma_1-\Gamma_3>0$ and $d_+=|{\bf d}_+|=d/\sqrt{2}$. The
ground-state configuration of the previous free energy has been already
discussed by several authors (see for example
\cite{Shekhtman,Thio,Papanicolaou}).  The spins order AF with an easy axis
in the $b$ direction, but with an additional small ferromagnetic (FM)
component along $c$, ${\bf S}_1^0=S(0,\cos\phi,\sin\phi)$ and ${\bf
S}_2^0=S(0,-\cos\phi,\sin\phi)$. The angle $\phi$ of the out-of-plane
canting of the spins is given by $ \phi_0=(1/2)\arctan
[2d_+/(2J-\a_c)]\approx d_+/2J$ and it is due to the DM interaction (see
also Fig.\ \ref{fig-canting} below). When this canting is taken into
account in the linearized equations of motion \pref{equations}, one can
easily see that the matrix for the transverse fluctuations in zero field
has the same structure of Eq.\ \pref{nofield}, with
\be
\lb{equiv}
\a_a=d_+\phi_0\approx d_+^2/(2J).
\ee
As a consequence, the effect of the DM interaction is twofold: it induces
the FM canting of the spins, and it reduces the AF coupling in the $a$
direction. The corresponding magnon gaps are, using Eq.\ \pref{gap-zf-anis}
and the equivalence \pref{equiv}:
\be
\lb{gap-zf}
\omega_a=m_a=zSd_+, \quad \omega_c=m_c=zS\sqrt{2\alpha_c J}.
\ee
Notice that the gap of the $a$ mode is proportional to $d$, while the gap
of the $c$ mode scales with the square-root of the parameter
$\a_c=\Gamma_1-\Gamma_3$. As a consequence, even though $\Gamma_i\sim
10^{-4} J$ and $d\sim 10^{-2} J$ the two gaps are of the same order of
magnitude in {\lco}.  When a finite magnetic field is applied the system
will evolve towards a new ground-state configuration. Following the
procedure described above the new magnon gaps can be determined. However,
we will not present these calculations here, because we shall describe in
detail in the next sections how these results can be obtained using the
NLSM approach, and how do they differ from the results \pref{hplla} and
\pref{hpllb}, that we shall refer to as 'conventional' in what follows.

\section{NLSM formulation for conventional anisotropies}

In this section we will show how the behavior of the spin gaps in the
presence of magnetic field can be easily derived within a NLSM description
of the low-energy physics of the spin model \pref{conv} and \pref{htot}.
We will first discuss the simple anisotropic model \pref{conv}, to show the
agreement with the results \pref{hplla} and \pref{hpllb} presented
above. Since the semiclassical approach is much more lengthly and less
transparent than the NLSM description, we shall adopt the latter to deal with
the more complicated case of the Hamiltonian \pref{htot}.

The derivation of the NLSM starting for the 2D Heisenberg
Hamiltonian has been extensively discussed in the
literature.\cite{sachdev} Here we just recall the main steps and
stress the origin of the mass terms due to the anisotropies in
Eq.\ \pref{conv}. First, we decompose the unit vector ${\bf
\Omega}_i= {\bf S}_i/S$ at site ${\bf r}_i$ into its
slowly-varying staggered and uniform components,
\be
{\bf \Omega}_i=\frac{{\bf S}_i(\tau)}{S}=e^{\rmi{\bf Q \cdot x}_i} {\bf n}({\bf
x}_i,\tau)+ a{\bf L}({\bf x}_i,\tau),
\ee
where ${\bf Q}=(\pi,\pi)$ and $a$ is the lattice parameter.  The constraint
${\bf \Omega}^2_i=1$ is enforced by ${\bf n}^2_i=1$ and $\bL_i\cdot
\bn_i=0$. Using this decomposition, the Heisenberg part of the Hamiltonian
(\ref{Hamiltonian}) has the standard form \cite{affleck}
\be
\lb{jj}
\LL_{HJ}=J\sum_{\langle i,j\rangle}{\bf S}_{i}\cdot{\bf S}_{j}=
\frac{JS^2}{2}\int \rmd^2 {\bf x} \left[ (\nabla\bn)^2+8\bL^2\right],
\ee
while the terms proportional to $\a_a,\a_c$ give rise to:\cite{sces05}
\be
\lb{gg}
\LL_{\a}=-\sum_{\langle
i,j\rangle}\a_aS_i^aS_j^a+\a_cS_i^cS_j^c=
\frac{S^2}{a^2}\int\rmd^2 {\bf x} \left[ 2\a_an^2_a+2\a_cn^2_c+
-\frac{1}{2}a^2\a_a(\nabla n_\a)^2 -\frac{1}{2}a^2 \a_c(\nabla
n_c)^2\right]. \ee
Since $\a_{a,c}\ll J$ we can neglect the correction induced by the
small anisotropies to the gradient of the transverse modes in Eq.\
\pref{jj}, and we can retain just the first two terms of
$\LL_\a$. Using a path-integral coherent states representation of the
spin states, which in addition to the previous contributions gives
rise to the (dynamical) Wess-Zumino term \cite{affleck}
\bea
\label{defl}
\LL_{WZ}=\frac{-iS}{a}\int \rmd^2{\bf x}\,\bL\cdot(\bn\times\dot\bn),\nn
\eea
we can obtain the partition function $Z=\int D\bn \delta(\bn^2-1)e^{-{\mathcal
S}}$, with the action ${\mathcal S}=\int \rmd \tau
[\LL_{HJ}+\LL_{\a}+\LL_{WZ}]$.  After integration of the $\bL$
fluctuations we obtain the following anisotropic
non-linear $\sigma$ model ($\beta=1/T$ and
$\int=\int_{0}^{\beta}\rmd\tau\int\rmd^{2}{\bf x}$):
\begin{equation}
\lb{nlsm-anis}
{\mathcal S}_0=\frac{1}{2gc}\int \left\{(\partial_{\tau}{\bf n})^{2}+
  c^2(\nabla{\bf n})^{2}+{ m_a^2} {n}_a^2+
m^2_c\; n_c^2
\right\}.
\end{equation}
The bare coupling constant $g$ and spin velocity $c$ are given by
$gc=8Ja^2$ and $c=2\sqrt{2}JSa$, and we defined $m^2_{a,c}=32JS^2\a_{a,c}$,
which corresponds to the result \pref{gap-zf-anis} above with $z=4$, as
appropriate in two dimensions. In generic $d$ dimensions the coefficients
$2\a_{a,c}$ in Eq.\ \pref{gg} are replaced by $z/2\a_{a,c}$ and one puts
$gc=4dJa^d=2zJa^d$, leading again to the definition \pref{gap-zf-anis} of
the masses. In the NLSM \pref{nlsm-anis} the spin stiffness is renormalized
by quantum fluctuations to
$\rho_{s}=c(1/Ng-\Lambda/4\pi)$,\cite{sachdev,CSY,CHN} where $\Lambda$ is a
cutoff for momentum integrals and $N=3$ is the number of spin components.
When the system orders we find the staggered magnetization at $\langle{\bf
n}\rangle=\sigma_0\hat{\bf x}_b$, because the orientation along $\hat{\bf
x}_a$ or $\hat{\bf x}_c$ would cost an energy $m_a$ or $m_c$, respectively.

It is worth noting that even though the NLSM \pref{nlsm-anis} contains
explicitly only the staggered spin component $\bn$, nonetheless the
saddle-point value of the uniform spin component $\bL$ determined before
integrating it out contains the residual information about the
ferromagnetically ordered spin component. This is evident when an external
magnetic field is applied on the system. In this case one can easily repeat
the previous calculations by taking into account that the saddle-point
value of the uniform magnetization $\bL$ acquires an additional
contribution proportional to $\bH$:\cite{Loss}
\be
\lb{lsp}
\bL=\frac{i}{8aSJ}(\bn\times\dot\bn)+\frac{1}{8aSJ}[\bH-\bn(\bn\cdot\bH)].
\ee
Observe that the first term is proportional to the time derivative of
$\bn$, so it averages to zero for the equlibrium configuration (indeed
no FM component is present in the ordinary AF phase). However, at
finite $\bH$ a non-vanishing average uniform component $\bL$ appears
in the field direction. After integration over $\bL$ the action
\pref{nlsm-anis} acquires additional terms proportional to $\bH$,
which can be recast into a shift of the time-derivative of $\bn$, as
expected from the spins precession around the applied field:
\be
\lb{nlsmh-anis}
{\cal S}({\bf H})=
{\cal S}_0(\partial_{\tau}{\bf n}\rightarrow\partial_{\tau}{\bf n}+
\rmi{\bH}\times{\bf n}) =
{\cal S}_0+\frac{1}{2gc}\int \left[ 2i
\bH\cdot (\bn \times \pd_\t \bn)-\bH^2+(\bH\cdot \bn)^2\right].
\ee
Observe that in the NLSM the constraint $\bn^2=1$ allows one to
rewrite the last two terms of Eq.\ \pref{nlsmh-anis} also as
$-\bH^2\bn_\perp^2$, where $\bn_\perp$ is the component of the order
parameter perpendicular to the field.

The non-linearity of the model \pref{nlsm-anis}-\pref{nlsmh-anis}
resides in the constraint $\bn^2=1$ for the staggered field. We will
implement it by means of a Lagrange multiplier $\lambda(\bx,\t)$, which
corresponds to add a term $\int i\lambda(\bn^2-1)$ to the action
\pref{nlsmh-anis}, and to perform an additional functional integration
over $\lambda$ in the partition function.\cite{CSY} In the saddle-point
approximation $\lambda$ can be taken as a constant $\lambda_0$, and one can
expand the field $\bn$ in terms of fluctuations around a given
equilibrium configuration $\bn_0$. Both the value of the Lagrange
multiplier and of the order parameter at each temperature will be
determined by minimizing the action with respect to them. The result
follows straighforwardly in the case of no magnetic field: assuming
$\bn_0=(0,\s_0,0)$, and integrating out in momentum space the
(transverse) Gaussian fluctuations around it, $\bn=(n_a,\s_0,n_c)$,
we obtain $Z=\exp(-\bar{\cal S})$ with:
\be \lb{sfluct} \bar {\cal S}=\frac{1}{2}Tr\log \hat G+ \frac{\b
\cA}{2gc}\left[m^2(\s_0^2-1)\right], \ee
where $\cA$ is the area of the 2D system, $m^2=i\lambda_02gc$ and the matrix
$\hat G^{-1}$ is given by:
\be
\lb{gm}
\hat G^{-1}=
\frac{1}{gc}\left( \begin{array}{cc}
\omega_n^2+c^2\bk^2+m_a^2+m^2 &  0 \\
 0       & \omega_n^2+c^2\bk^2+m_c^2+m^2
\end{array} \right),
\ee
where $\omega_n=2\pi nT$ and $\bq$ are the Matsubara frequencies and the momenta,
respectively, and the trace in Eq.\ \pref{sfluct} is over $\omega_m,\bq$ and
the matrix indexes. By minimizing the action \pref{sfluct} we obtain two
equations:
\bea
m^2\s_0&=&0,\nn\\
\lb{sce}
\s_0^2&=&1-NI_\perp,
\eea
where $I_\perp=(1/2)(I_a+I_c)$ accounts for the transverse
fluctuations, with
\be \lb{iperp} I_{a,c}=\frac{1}{\b \cA}\sum_{\bk,\omega_n}
G_{a,c}(\bk,\omega_n)= \frac{1}{\b \cA}\sum_{\bk,\omega_n} \langle|
n_{a,c}(\bk,\omega_n)|^2\rangle, \ee
where $G_{a,c}=\langle n^2_{a,c}\rangle$ is the Green function for the
$a,c$ mode. From Eq.\ \pref{sce} we see that two regimes are
possible:\cite{sachdev,CSY} (i) $\s_0=0, m\neq 0$, which corresponds to the
paramegnetic phase. Here $m^2$ plays the role of the inverse correlation
length, defined by the second of Eqs.\ \pref{sce} for $\s_0=0$,
i.e. $1=NI_\perp(m^2)$; (ii) $m=0,\s_0\neq 0$ which is the ordered phase,
where the order parameter is an increasing function of temperature below
$T_N$, defined as the temperature at which the mass first vanishes,
i.e. $1=NI_\perp(0)$. Observe that in the 2D case the functions $I_{a,c}$
can be evaluated analytically and are given by:
\be
\lb{i2d}
I_{a,c}=\frac{gT}{2\pi c}\ln\left\{
\frac{\sinh(c\Lambda/2T)}{\sinh(\omega_{a,c}/2T)}\right\}.
\ee
As far as the magnon gaps are concerned, they are defined as the
poles of the spectral function at zero momentum:
\be
\lb{spectralf}
\cA_{a,c}(\omega)=-\frac{1}{\pi}{\rm Im} G_{a,c}(i\omega_n\ra
\omega+i0^+,\bk=0)=\frac{1}{2m_{a,c}}\left[ \d(\omega-m_{a,c})-\d(\omega+m_{a,c})
 \right],
\ee
where the last equality only holds in the case of a matrix for the
transverse fluctuations having the diagonal structure of Eq.\
\pref{gm}. Thus, one can identify $m_{a,c}$ as the magnon gaps at zero field.

\subsection{Transverse field}
Let us analyze how the previous results are modified in the presence
of a finite magnetic field. We first consider the case of a transerse
field, for example $\bH\pll a$. Eq.\ \pref{nlsmh-anis} then reads:
$$ 
{\cal S}={\cal S}_0+\frac{1}{2gc}\int \left[ 2i H (n_b \pd_\t n_c-n_c
\pd_\t n_b) -H^2+H^2 n_a^2+i\lambda 2gc(\bn^2-1)\right].
$$
As a consequence, using again $\bn=(n_a,\s_0,n_c)$, the only
modification to the previous set of equations is the replacement of $m_a^2$
with $m_a^2+H^2$ in the Eq.\ \pref{gm} defining the transverse
fluctuations. Thus, one recovers the same phase transition as before
(with negligible quantitative corrections to the $T_N$ and the $\s_0$
value). As far as the magnon gaps are concerned, we see that the $c$
mode is unchanged, while it occurs a shift of the mass of the $a$ mode,
leading to two poles at:
$$
\omega_a^2=m_a^2+H^2, \quad \omega_c^2=m_c^2,
$$
corresponding to the result \pref{hplla} that we derived above.

\subsection{Longitudinal field}
In the case instead of a longitudinal field, $\bH\pll b$, Eq.\
\pref{nlsmh-anis} reads:
$$
{\cal S}={\cal S}_0+\frac{1}{2gc}\int \left[ 2i H (n_a \pd_\t n_c-n_c \pd_\t
n_a) -H^2(n_a^2+n_c^2)+i\lambda 2gc(\bn^2-1)\right].
$$
Thus, when $\bn=(n_a,\s_0,n_c)$ we find that the equivalent of the
inverse matrix \pref{gm} acquires off-diagonal terms proportional to
the applied field:
$$
\hat G^{-1}=
\frac{1}{gc}\left( \begin{array}{cc}
\omega_n^2+c^2\bk^2+m_a^2-H^2+m^2 &  2\omega H \\
 -2\omega H      & \omega_n^2+c^2\bk^2+m_c^2-H^2+m^2
\end{array} \right).
$$
As a consequence, the matrix $\hat G$ which defines the transverse
fluctuations reads:
\be
\lb{gmb}
\hat{ G}=\frac{(gc)}{{\rm det} \, \hat G}
\left( \begin{array}{cc}
\omega_n^2+\varepsilon^2_c(\bk)-H^2 &  -2\omega H \\
 2\omega H      & \omega_n^2+\varepsilon^2_a(\bk)-H^2
\end{array} \right),
\ee
where $\varepsilon^2_{a,c}(\bk)=c^2\bk^2 +m_{a,c}^2$. Due to this structure,
the poles of the spectral functions for the $a,c$ modes are the zeros
of the determinant of the $\hat G$ matrix at $\bk=0$ and $i\omega_n\ra
\omega+i0^+$:
$$
(-\omega^2+\varepsilon^2_a({\bf 0})-H^2)(-\omega^2+\varepsilon^2_c({\bf
0})-H^2)-4\omega^2H^2=0, 
$$
and correspond to the two solutions \pref{hpllb} determined above
using semiclassical spin-wave theory. Observe that in principle both
solutions appear in the spectral function of the $a$ or $c$
mode. However, the spectral weight associated to the two solutions
$\omega_{a,c}$ differs in the two cases. For example, for the $a$ mode we
have:
\be \lb{ca} \cA_a(\omega>0)=
\left[\frac{Z_a}{2\omega_a}\d(\omega-\omega_a)+\frac{Z_c}{2\omega_c}\d(\omega-\omega_c)\right],
\ee
where the residua at the poles are (see also Eq.\ \pref{zq} below)
$Z_{a,c}=\pm(-\omega^2_{a,c}+m^2_c-H^2)/(\omega_c^2-\omega_a^2)$. Since
$Z_a/\omega_a\gg Z_c/\omega_c$ one can conclude that the spectral function of
the $a$ mode is dominated by the pole at $\omega_a$, and conversely for
the $c$ mode, confirming the identification of the two function
\pref{hpllb} as the correct magnon gaps in an applied longitudinal
field.

\subsection{In-plane field}

\begin{figure}[htb]
\includegraphics[angle=-90,scale=0.4]{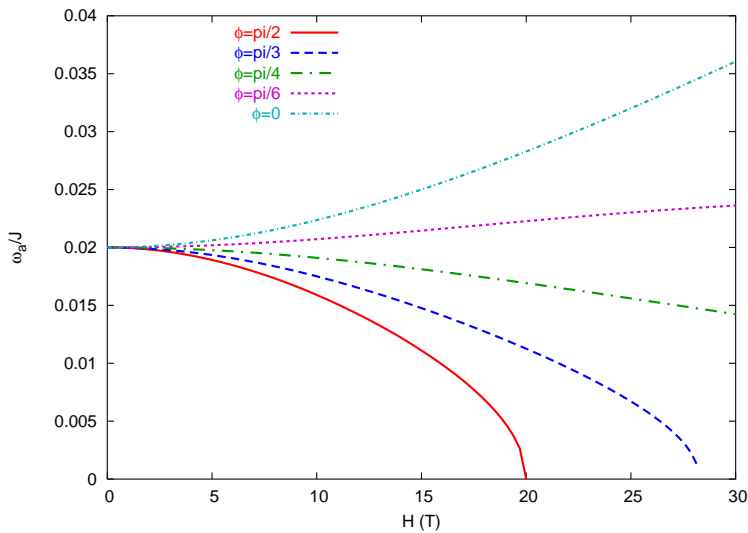}
\caption{(Color online) Field dependence of the in-plane magnon gap for an
  in-plane field applied at an arbitrary angle $\phi$ with the $a$ axis,
  see Eq.\ \pref{inplane}.  The case $\phi=0$ corresponds to the
  transverse-field case analyzed in Sec. IV-A, Eq.\ \pref{hplla}, while the
  case $\phi=\pi/2$ corresponds to a purely longitudinal field, Eq.\
  \pref{hpllb}, which has a spin-flop transition at $H=m_a$. The behavior
  of the gap above the spin-flop transition is described in Sec. V-A below.}
\label{fig-inplane}
\end{figure}

Finally, for the sake of completeness we analyze the case when the field is
applied in the plane forming an arbitrary angle $\phi$ with the $a$
axis. Thus, the field has both a longitudinal ($H\sin\phi$) and a
transverse ($H\cos\phi$) component, and we expect an intermediate behavior
between the two cases analyzed above. Following the same line of
calculculations described in the previous subsections, we obtain:
\bea 
\omega_a^2&=&\frac{m_a^2+m_c^2}{2}+\frac{H^2}{2}A\nonumber\\
\lb{inplane}
&-&\sqrt{\left(\frac{m_a^2-m_c^2}{2}\right)^2+\frac{H^2}{2}A(m_a^2+m_c^2)+
m_a^2H^2\sin^2{\phi}-m_c^2H^2\cos{2\phi}+\frac{H^4}{4}[A^2+B]},
\eea
where
\bea
A=3\sin^2{\phi}+\cos{2\phi},\nonumber\\
B=4\cos{2\phi}\sin^2{\phi}.\nonumber
\eea
For the $c$ mode we obtain an analogous expression, with a plus sign in
front of the square-root term in Eq.\ \pref{inplane}. Thus, Eq.\
\pref{inplane} reduces to the results \pref{hplla} and \pref{hpllb} when
$\phi=0$ and $\phi=\pi/2$, respectively. The behavior of $\omega_a$ for
various values of the angle $\phi$ as a function of the field strength $H$
is plotted in Fig.\ \ref{fig-inplane}. At $\phi=\pi/2$ the expression
\pref{inplane} (and then also Eq.\ \pref{hpllb}) is vanishing at a field
$H_c=m_a$. Indeed, as we discuss in detail in Sec. V-A, at this critical
field the spins perform an in-plane spin-flop transition to orient
perpendicularly to the magnetic field. When $\phi$ deviates from $\pi/2$
the longitudinal field component decreases and the spin-flop transition
moves to a higher value of the field.  Accordingly, the field dependence of
the gap changes continuosly, going from a decreasing function to an
increasing one, recovering at $\phi=0$ the increasing behavior
dictated by Eq.\ \pref{hplla}, characteristic of a purely transverse
field. From Fig.\ \ref{fig-inplane} it is also clear that the two extreme
cases are also the ones where the largest deviation of the gap from the
zero-field value can be observed.

\section{NLSM with interlayer coupling and DM interaction}
\begin{figure}[htb]
\includegraphics[angle=0,scale=0.5]{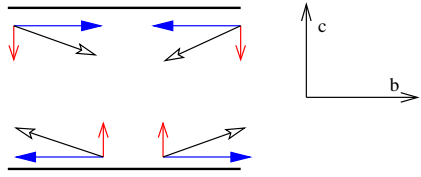}
\caption{(Color online) Spin configuration at zero applied magnetic field
  in {\lco}. The arrows with a solid tip represent the staggered spin
  components $(-1)^m e^{i\bQ \cdot {\bf r}_i}\bn_i^m$, the arrows with a
  two-lines tip represent the uniform spin components $\bL_i^m$, the
  arrows with an open tip the full spins $\bS_i^m$.}
\label{fig-canting}
\end{figure}

Once we have established the equivalence between the semiclassical
approach and the NLSM derivation of the spin-wave gaps for an
ordinary easy-axis antiferromagnet, we can discuss the more
general case of the model \pref{htot} where also interlayer
coupling and DM interactions are present. The NLSM for this system
has been already derived in Refs.\onlinecite{MLVC,sces05,Papanicolaou},
where it has been shown that in the absence of magnetic field a 3D
analogous of Eq.\ \pref{nlsm-anis} holds:
\be
\lb{nlsm}
{\cal S}_0=\frac{1}{2gc}\sum_m\int \left\{
(\pd_\t\bn_m)^2+c^2(\nb\bn_m)^2+\eta(\bn_m-\bn_{m+1})^2+(D_+ n_m^a)^2+\Gamma_c
(n_m^c)^2\right\},
\ee
where $m$ in the layer index, $\bD_+=D_+\hat\bx_a$, $D_+=4Sd_+$,
$\Gamma_c=32JS^2(\Gamma_1-\Gamma_3)$  and $\eta=2JJ_\perp$. In this
notation the in-plane and out-of-plane modes have masses:
\be
m_a=D_+=4Sd_+, \quad m_c=\sqrt{\Gamma_c}=4S\sqrt{2J(\Gamma_1-\Gamma_3)},
\ee
in agreement with the results \pref{gap-zf} above. Moreover, the uniform
magnetization of each layer acquires an additional contribution
proportional to the DM vector $\bD_+$ with respect to Eq.\ \pref{lsp} :
\be
\lb{lspdm}
\bL_m=\frac{i}{8aSJ}(\bn_m\times\dot\bn_m)+
\frac{1}{8aSJ}[(-1)^m\bn_m\times \bD_+ +\bH-\bn_m(\bn_m\cdot\bH)],
\ee
where the oscillating factor $(-1)^m$ in the term proportional to
$\bD_+$ accounts for the effect of the tilting of the ochtaedra on
neighboring planes, as discussed below Eq.\ \pref{htot}. As one can
easily see applying the saddle-point approach described in the previous
Section to the action \pref{nlsm}, at $\bH=0$ the system orders AF
below $T_N$ in a 3D staggered configuration, with $\bn_m$ along $b$ in
each layer. Moreover, due to the oscillating factor $(-1)^m$ in Eq.\
\pref{lspdm}, the spins in each layer acquire a FM components $\bL_m$,
with the vectors $\bL_m$ ordered AF in neighboring layers, see Fig.\
\ref{fig-canting}. 
\be \lb{unif} \bH=0, \quad \langle \bn_m\rangle =\s_0\hat\bx_b,
\quad \langle \bL_m\rangle=(-1)^m \frac{\langle \bn_m\rangle
D_+}{8aSJ}=(-1)^m\frac{\s_0D_+}{8aSJ}. \ee
The additional term in $\bD_+$ in Eq.\ \pref{lspdm} translates in an
additional coupling between the order parameter and $\bH$ when the full
NLSM action at finite magnetic field is computed:
\be
\lb{nlsmh}
{\cal S}({\bf B})=
{\cal S}_0+\frac{1}{2gc}\sum_m\int \left[ 2i
\bH\cdot (\bn_m \times \pd_\t \bn_m)-\bH^2+(\bH\cdot \bn_m)^2
-(-1)^m2\bH\cdot (\bn_m\times \bD_+)\right].
\ee
As it has been discussed in
Refs. \onlinecite{MLVC,Thio,Thio90,Papanicolaou,Raman}, the {\em effective}
staggered field acting on the AF order parameter due to the presence of the DM
interaction makes the system an {\em unconventional} easy-axis AF. As far
as the spin-waves gaps are concerned, the results of the previous Sections
apply only in some specific regimes, as we shall analyze below. It is worth
noting that the last three terms in Eq.\ \pref{nlsmh} are proportional to:
\be \lb{heff} -\bH\cdot(\bL^H_m+\bL^{DM}_m), \ee
where $\bL^H_m$ and $\bL^{DM}_m$ are the contributions to $\bL_m$ in Eq.\
\pref{lspdm} proportional to the magnetic field and to the DM term,
respectively. As a consequence, the ground-state of the action
\pref{nlsmh} will be determined by the competition between the
energetic cost of the bare action ${\cal S}_0$ and the tendency of the
system to maximize the uniform spin components in the field direction,
to gain energy from the term \pref{heff}. Even though part of the
ground-state phenomenology has been already described in
Ref. \onlinecite{Thio,Thio90,Papanicolaou,MLVC,Raman}, here we will derive
these results within the general language of the saddle-point
approximation for the NLSM, by focusing on the magnon-gaps
behavior,\cite{Papanicolaou} that will be compared with the
expectation for an ordinary easy-axis AF, described in the previous
Section. Then we shall also compute the effect of quantum and thermal
corrections, which allows us to investigate the $(H,T_N)$ phase
diagram in the various case.

\subsection{ $\bH$ paralell to $b$}
\begin{figure}[htb]
\includegraphics[angle=-90,scale=0.4]{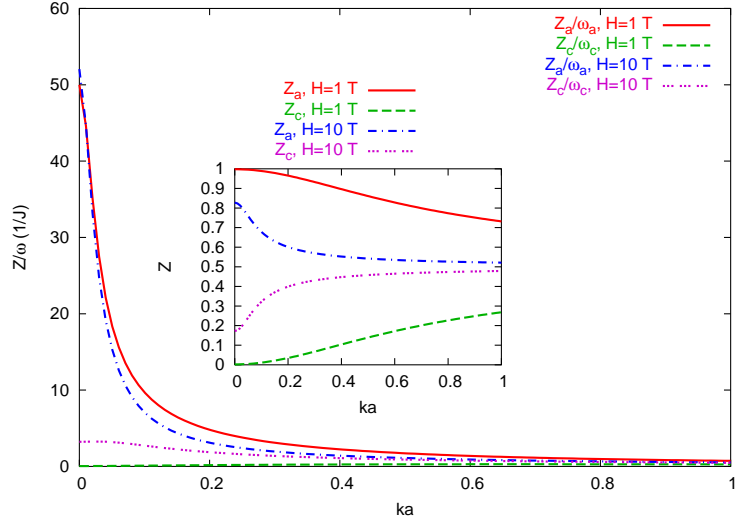}
\caption{(Color online) Momentum dependence of the spectral weight
  corresponding to the poles at $\omega_a(\bk,k_\perp)$ or
  $\omega_c(\bk,k_\perp)$ in Eq.\ \pref{iahpllb}. Here $k_\perp=0$,
  $m_a=0.02 J, m_c=0.05 J, c=J$, and we used units such that H=1 T
  corresponds to $10^{-3}J$. Inset: momentum dependence of $Z_a, Z_c$
  defined in Eq. \pref{zq}. Observe that even thought $Z_a\approx Z_c$ at
  $H=10$ T as $ka \sim 0.5$ the contribution at $k=0$ is always predominant
  in the momentum sum \pref{iahpllb}, due to the fact that the $a$ gap
  softens as the field strenght increases.}
\label{polesb}
\end{figure}

\begin{figure}[htb]
\includegraphics[angle=0,scale=0.4]{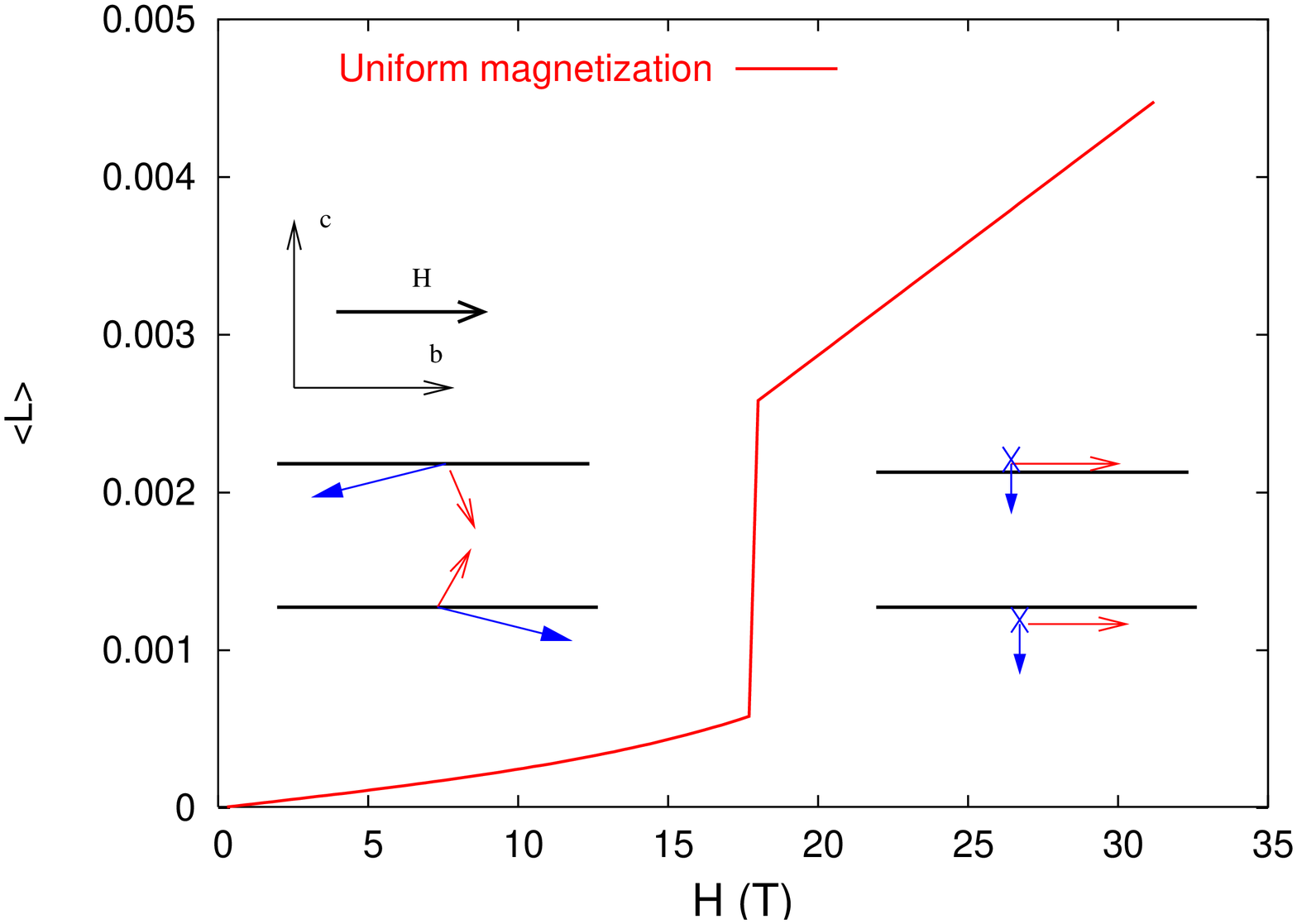}
\caption{(Color online) Magnetic-field dependence at of the spins ground
  state and of the $T=0$ average uniform magnetization in the field
  direction for a field along the $b$ axis. At finite magnetic field the
  order-parameter $\bn$ rotates in the $bc$ plane. The spin configuration
  is determined as usual by $\bS_i/S=(-1)^me^{i\bQ \cdot {\bf
  r}_i}\bn_i^m+\bL_i^m$, where the staggered and uniform components are
  indicated by arrows with different tips as in Fig.\ \ref{fig-canting}:
  the arrows with a two-lines tip represent the uniform components $\bL_m$
  and the arrows with a solid tip the staggered components $(-1)^me^{i\bQ
  \cdot {\bf r}_i}\bn_i^m$.  At $H<H_c^{(1)}\approx 18$ T (left in
  the figure) $\bn_m$ is defined in Eq.\ \pref{solb1} and $\bL_m$ in Eq.\
  \pref{bunif1}, so that the spins have both the staggered and uniform
  component in the $bc$ plane. The uniform components are due both to the
  magnetic field and to the DM interaction, see Eq.\ \pref{bunif1}, and the
  sum of $\bL_M+\bL_{m+1}$ is alligned along the field. Above the spin-flop
  transition (right in the figure) the staggered components $\bn_m$ lie in
  the $ac$ plane see Eq.\ \pref{solb2} (so that the part along $a$ is
  orthogonal to the plane of the figure and indicated by a cross) and the
  uniform components $\bL_m$ point along $b$, see Eq.\ \pref{bunif2}. The
  order-parameter values used to evaluate the $\langle \bL\rangle$ are the
  same reported in Fig.\ \ref{phase-dia}. Observe that since in our
  approximation the critical field $H_c^{(1)}$ is independent of
  temperature, as well as $\s_c$, the uniform magnetization is temperature
  independent below $T_N$. }
\label{uniformb}
\end{figure}

When the field is in the $b$ direction the last term in Eq.\ \pref{nlsmh}
generates a staggered field along the $c$ direction which leads to a
rotation of the order parameter in the $bc$
plane:\cite{Thio90,Papanicolaou,Raman}
$$
{\cal S}({\bf B})=
{\cal S}_0+\frac{1}{2gc}\sum_m\int \left[
  2iH(n_m^a\pd_\t n_m^c-n_m^c\pd_\t n_m^a)-H^2[(n_m^c)^2+(n_m^a)^2]
+2h^mn_m^c+i\lambda_m 2gc(\bn_m^2-1) \right], 
$$
where we put $h^m=(-1)^m HD_+$ and we introduced explicitly also a set of
Lagrange multipliers $\lambda_m$ which enforce the constraint $\bn_m^2=1$ in
each layer.  Due to the anomalous coupling, a finite component in the $c$
direction can arise. For a generic configuration $\bn_m=(0,\s^b_m,\s^c_m)$
we find the ground-state action ($m^2_m=2i\lambda_m gc$):
$$
\bar{\cal S}_{cl}=\frac{\b\cA}{2gc}\sum_m
\eta(\s_m^b-\s_{m+1}^b)^2+\eta(\s_m^c-\s_{m+1}^c)^2
-2h^m\s_m^c+(m_c^2-H^2)(\s_m^c)^2+m^2_m[(\s_m^b)^2+(\s_m^c)^2-1].
$$
The saddle-point equations then read:
\bea
\eta(2\s_m^b-\s_{m+1}^b-\s_{m-1}^b)+m^2_m\s_m^b=0,\nn\\
(m_c^2-H^2)\s_m^c+h^m+\eta(2\s_m^c-\s_{m+1}^c-\s_{m-1}^c)+m^2\s_m^c=0,\nn\\
(\s_m^b)^2+(\s_m^c)^2=1.\nn \eea
At low field, one can easily check that the classical configuration is
given by an order parameter with a uniform $\s_b$ component in neighboring
layers and an oscillating $\s_c$ component (which corresponds to the $c$
components of the spins coming from the $(-1)^m e^{i\bQ \cdot {\bf
r}_i}\bn^m_i$ term ordered ferromagnetically in neighboring planes, see
Fig.\ \ref{uniformb}):
\bea
\bar{\cal S}_{cl}=-\frac{\b\cA}{2gc}|\s_c| h, &\quad&
m^2_m=m^2,  \nn\\
\lb{solb1}
\s_b^m=\s_b=1-\s_c^2,
&\quad& \s_m^c=(-1)^m\s_c=(-1)^m\frac{-HD_+}{m_c^2+4\eta-H^2}.
\eea
In this configuration the $\bL_m$ vectors are given by:
\bea
\bL_m^{H}&=&\frac{H}{8aSJ}\left[ \s_c^2\hat\bx_b-(-1)^m\s_c\s_b\hat\bx_c
  \right],\nn\\
\lb{bunif1}
\bL_m^{DM}&=&\frac{D_+}{8aSJ}\left[ |\s_c|\hat\bx_b+(-1)^m\s_b\hat\bx_c
  \right],
\eea
so that the average (i.e. summed over neighboring layers) uniform
magnetization is along $b$ and given by $\langle \bL \rangle
=(1/8aSJ)[D_+|\s_c|+H\s_c^2] \hat\bx_b$. Thus, the oscillating character of
$\s_m^c$ allows for the DM-induced magnetization to allign in the direction
of the field, see Fig.\ \ref{uniformb}. After inclusion of the transverse
$a,c$ fluctuations, the order-parameter equations read:
\bea
\s_b^2&=&1-\s_c^2-NI_\perp(m=0),\nn\\
\lb{t<tn1}
\s_c&=&-\frac{HD_+}{m_c^2+4\eta-H^2}, \quad T<T_N,
\eea
below $T_N$ and
\bea
\s_b&=&0, \Ra 1=\s_c^2+NI_\perp(m),\nn\\
\lb{t>tn1}
\s_c&=&-\frac{HD_+}{m_c^2+4\eta-H^2+m^2}, \quad T>T_N,
\eea
above $T_N$, where $I_\perp$ is computed using the matrix \pref{gmb}
discussed above for the case of longitudinal field. Moreover, since
the system is now 3D, we have that the energy dispersion of the transverse
modes is
\be
\lb{defeq}
\varepsilon^2_{a,c}(\bk,k_\perp)=c^2\bk^2+2\eta(1-\cos k_\perp
d)+m_{a,c}^2,
\ee
where $d$ is the interlayer sapcing and an additional integration over
out-of-plane momentum $k_\perp$ must be included in computing
$I_{a,c}$. Thus, taking into account the non-diagonal character of the
fluctuations matrix \pref{gmb}, we have for example for the $a$ mode:
\be
\lb{iahpllb}
I_{a}=
\frac{1}{\b V}\sum_{\omega_n,\bk,k_\perp}
\langle |n_{a}(\omega_n,\bk,k_\perp)|^2\rangle=
\frac{1}{V}\sum_{\bk,k_\perp}
\frac{Z_a(\bk,k_\perp)}{2\omega_a(\bk,k_\perp)}\coth \frac{\b
  \omega_a(\bk,k_\perp)}{2} +
\frac{Z_c(\bk,k_\perp)}{2\omega_c(\bk,k_\perp)}\coth \frac{\b
  \omega_c(\bk,k_\perp)}{2},
\ee
where $V$ is the 3D volume of the system.  Here $\omega_{a,c}$ are the
generalization of Eq.\ \pref{hpllb} at finite momentum:
\be
\lb{oack}
\omega^2_{a,c}(\bk,k_\perp)=\frac{\varepsilon_c^2+\varepsilon_a^2}{2}+H^2\pm
\sqrt{\left(\frac{\varepsilon_c^2-\varepsilon_a^2}{2}\right)^2+
4H^2\left(\frac{\varepsilon_c^2+\varepsilon_a^2}{2}\right)},
\ee
where the explicit dependence of $\varepsilon_{a,c}$ on momenta has been
omitted.  Analogously, the spectral weights $Z_{a,c}$ of the two poles are
given by:
\be \lb{zq} Z_{a,c}(\bk,k_\perp)=\pm
\frac{-\omega_{a,c}^2+\varepsilon_c^2(\bk,k_\perp)
-H^2}{\omega_c^2(\bk,k_\perp)-\omega_a^2(\bk,k_\perp)}, \ee
and are plotted in the inset of Fig.\ \ref{polesb} for $k_\perp=0$ as a
function of $ka$. Note that since in all the formulas the magnetic field is
measured in units of $g_s\mu_B\approx 0.1$ meV, and since typical values of
$J$ are of the order of 130 meV, we used for simplicity the equivalence
$H=1$ T=$10^{-3}J$. 
%
%
Due to the thermal factors, the main contribution in
the momentum integration in Eq.\ \pref{iahpllb} comes from $\bk=k_\perp=0$,
where also the factors $Z_\a/\omega_a$ have the largest value, see Fig.\
\ref{polesb}. As a consequence, we can safely approximate the momentum
dependence of Eq.\ \pref{oack} with
\be
\lb{oappr}
\omega^2_{a,c}(\bk,k_\perp)\approx \omega^2_{a,c}+c^2\bk^2+2\eta(1-\cos k_\perp d),
\ee
where $\omega_{a,c}$ are the magnon gaps given in Eq.\ \pref{hpllb}, and we can
neglect the momentum dependence of $Z_{a,c}$ in Eq.\ \pref{iahpllb}. We
thus obtain below $T_N$:
\be \lb{ia} I_{a}=Z_a(0) I_{3D}(\omega_a)+Z_c(0) I_{3D}(\omega_c), \ee
where $I_{3D}(M)$ is the extension to the layered
3D case of the integral \pref{i2d}:
\be
\lb{i3d}
I_{3D}(M)=
\frac{gT}{2\pi c}\int_{-\pi}^\pi \frac{dz}{2\pi}
\ln\left\{
\frac{\sinh(c\Lambda/2T)}{\sinh \sqrt{M^2+2\eta(1-\cos z)}/2T}\right\}.
\ee
Above $T_N$ we simply substitute $\omega_{a,c}^2 \ra \omega_{a,c}^2+m^2$ in Eq.\
\pref{ia}, where as usual $m^2$ plays the role of the inverse correlation
length, to be obtained solving the self-consistency equation \pref{t>tn1}.
As far as the $c$ fluctuations are concerned, the previous result is
clearly reversed, with a larger contribution coming from the pole at
$\omega_c$, since now $Z_{a,c}(0)=\mp (-\omega_{a,c}^2+m_a^2-H^2)/(\omega_c^2-\omega_a^2)$.

As the field strength increases we see that, according to Eq.\
\pref{hpllb}, the smaller gap $\omega_a$ decreases,
and vanishes at the critical field:
\be
\lb{hc1}
H_c^{(1)}=m_a=D_+.
\ee
Indeed, at $H_c^{(1)}$ we have a spin-flop transition: the in-plane component
of the order parameter rotate from the $b$ to the $a$ direction, so that the
classical configuration becomes:
\be
\lb{solb2}
\bn_m=(\s_a,0,(-1)^m\s_c).
\ee
The uniform magnetization changes correspondingly:
\be
\lb{bunif2}
\bL_m^{H}=\frac{H}{8aSJ}\hat\bx_b, \quad  \bL_m^{DM}=\frac{D_+}{8aSJ}|\s_c|
\hat\bx_b,
\ee
and jumps discontinously at the spin-flop transition by a quantity
$H[1-\s_c(H_c^{(1)})]/8aSJ$, see Fig.\ \ref{uniformb}. 

\begin{figure}[htb]
\includegraphics[angle=-90,scale=0.4]{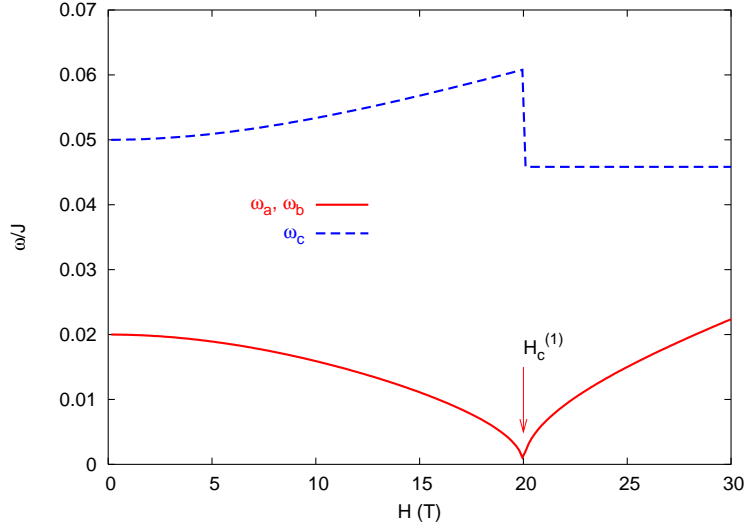}
\caption{(Color online) Field dependence of the magnon gaps at $T=0$ for a
  field applied along the $b$ axis. We used $m_a=0.02 J, m_c=0.05 J$, and
  units such that H=1 T corresponds to $10^{-3}J$, so that the first
  critical field $H_c^{(1)}=D_+=m_a=20$ T. Below $H_c^{(1)}$ the in-plane
  and out-of-plane mode are correspond to $a$ and $c$ component
  fluctuations, whose gaps are given by $\omega_a$ and $\omega_c$ in Eq.\
  \pref{hpllb}, respectively. Above the critical field the in-plane mode
  corresponds to fluctuations of the $b$ component, and the magnon gaps are
  defined by Eq.\ \pref{h>hc1}.}
\label{gap-baxis}
\end{figure}

By adding $b$ and $c$ fluctuations around this ground-state
solution we obtain the new saddle-point equations:
\bea
\s_a^2&=&1-\s_c^2-N\tilde{I}_\perp(m=0),\nn\\
\lb{t<tn2}
\s_c&=&-\frac{HD_+}{m_c^2+4\eta-m_a^2}, \quad T<T_N,
\eea
and
\bea
\s_a&=&0, \Ra 1=\s_c^2+N\tilde I_\perp(m),\nn\\
\lb{t>tn2}
\s_c&=&-\frac{HD_+}{m_c^2+4\eta-m_a^2+m^2}, \quad T>T_N,
\eea
where $\tilde I_\perp$ accounts for the $b,c$ fluctuations described by the
inverse matrix:
\be
\lb{gm-hb}
\hat G^{-1}=
\left( \begin{array}{cc}
\omega^2(\bq)+H^2-m_a^2+m^2 &  0 \\
 0       & \omega^2(\bq)+m_c^2-m_a^2+m^2
\end{array} \right),
\ee
where $\omega^2(\bq)=\omega_n^2+c^2\bk^2+2\eta(1-\cos k_\perp d)$.  As far as the
magnon gaps are concerned, we see that above the transition the field in
the $b$ direction acts as a {\em transverse} field, since the direction of
the magnetization has changed. Once again, the behavior of the magnon gaps
can be easily read from the Green's function matrix \pref{gm-hb}. We find
that the in-plane mode corresponds now to a fluctuation of the $b$
component, with a field-dependent mass, while the out-of-plane mode does
not depend on the field but should be rescaled with respect to the $m_a$
gap:
\be
\lb{h>hc1}
\omega^2_{in}\equiv\omega_b^2=\sqrt{H^2-m_a^2}, \quad \omega_c^2=\sqrt{m_c^2-m_a^2}.
\ee
The resulting field dependence of the magnon gaps is reported in Fig.\
\ref{gap-baxis}.

\begin{figure}[htb]
\includegraphics[angle=-90,scale=0.4]{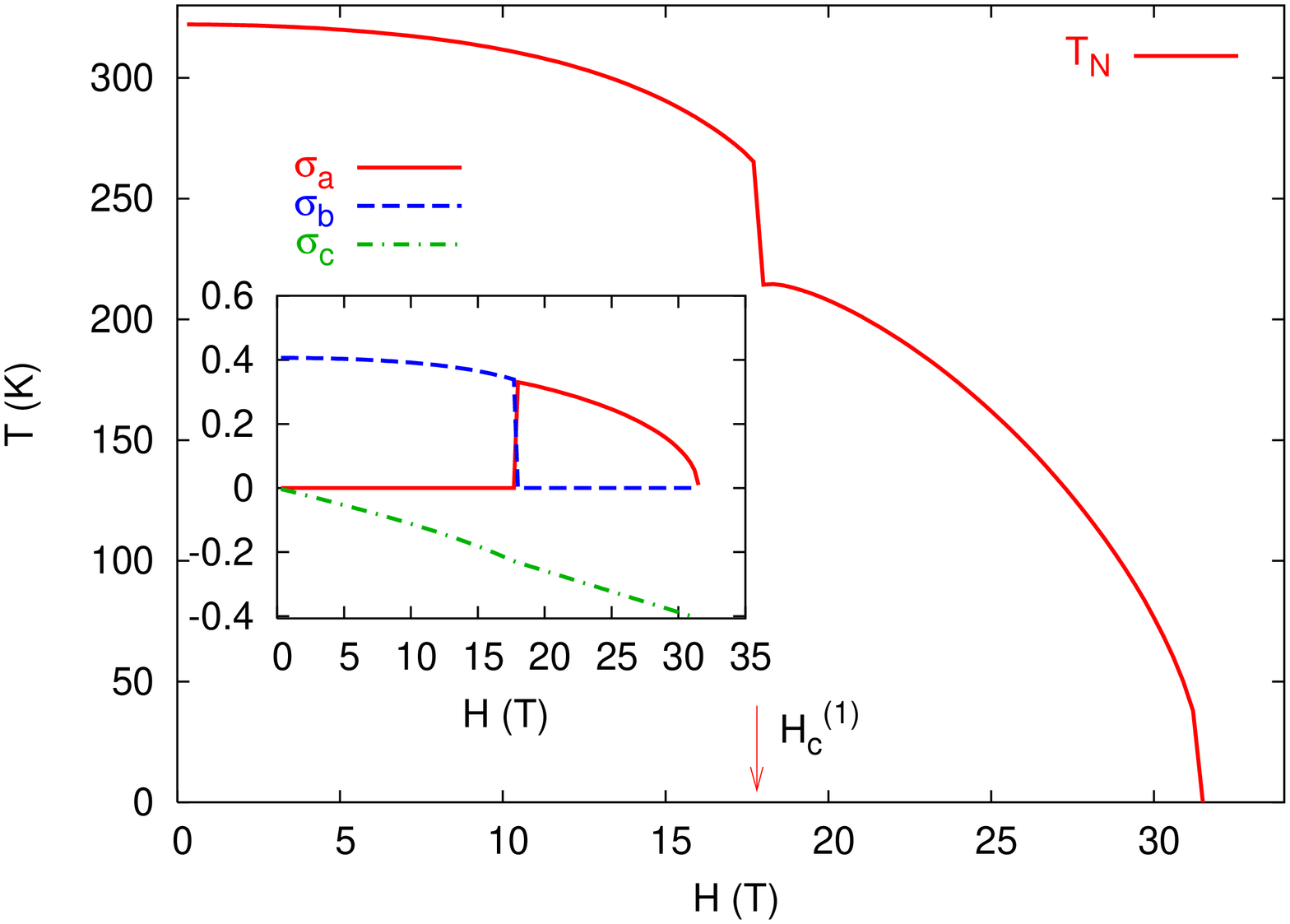}
\caption{(Color online) Phase diagram of the system for $\bH\pll b$. The
  parameter values are chosen to reproduce the data of {\lco},
  i.e. $D_+=m_a=0.0116 J$, $m_c=0.034 J$, $J=130$ meV, $\rho_s=0.07 J$ and
  $c=1.3J$ \cite{MLVC}. For the interlayer coupling we used $\eta=8\times
  10^{-5} J^2$, which correctly reproduce the spin-flop transition measured
  in Ref.\ \onlinecite{Gozar} for $\bH \pll c$, see end of Sec. V-B. The magnetic
  field is converted in energy trough $H\ra g_s\mu_B H$, where $\mu_B$ is
  the Bohr magneton and $g_s=2.1$ is the gyromagnetic ratio in the $b$
  direction. Inset: field evolution of the staggered order parameter 
  components at $T=0$ in the various phases.}
\label{phase-dia}
\end{figure}

By further increasing $H$ one will finally reaches a second critical field
above which the
transverse in-plane component of the staggered order parameter vanishes,
$\sigma_a=0$, but the $\sigma_c$ component is still nonzero, see Fig.\
\ref{phase-dia}. Since by definition $|\bn|<1$, we can see from 
Eq.\ \pref{t<tn2} that the solution \pref{solb2} is valid only 
when $\sigma_c\leq 1$, so at fields lower than \cite{Thio90,Papanicolaou}:
\be
\lb{hc2naive}
H_c^{(2,naive)}=(m^2_c+4\eta-m_a^2)/D_+.
\ee
However, the estimate \pref{hc2naive} does not take into account quantum
fluctuations, which reduce the $T=0$ value of the in-plane order parameter
according to Eq.\ \pref{t<tn1}. Since the transverse fluctuations $I_\perp$
do not depend strongly on the magnetic field, one can see that the second
critical field, defined as the field at which $\s_a(T=0)=0$ in Eq.\
\pref{t<tn2}, is given approximately by:
\be
\lb{hc2}
H_c^{(2)}\simeq \frac{m^2_c+4\eta-m_a^2}{D_+} \s_0(H=0,T=0),
\ee
where $\s_0(H=0)=\sqrt{1-NI_\perp(T=0,H=0)}$ is the magnetization of
the system measured at $T=0$ along the $b$ axis without external
magnetic field. Observe that quantum corrections can indeed reduce
considerably the second critical field. 
%
%
In Fig.\ \ref{phase-dia} we report as an example the phase diagram for
{\lco} of the transition temperature $T_N$ vs $H$ obtained in the
saddle-point approximation. By estimating the parameter values from the
Raman measurements of the magnon gaps in Ref. \onlinecite{Gozar} (see also
\onlinecite{theo-exp}), we choose $m_a=0.0116J$, $m_c=0.034J$, $\eta=8\times
10^{-5} J^2$, with $J=130$ meV and $g_s=2.1$, as appropriate for the $b$
direction. For the stiffness and spin-wave velocity we use $\rho_s=0.07J$
and $c=1.3J$ respectively, which are not far from the standard values
quoted in the literature \cite{CHN,CSY} and have been shown to be
appropriate to reproduce (in the same approximation) the
uniform-susceptibility data \cite{MLVC}. In the inset we also report the
$T=0$ value of the order-parameter components as a function of the
field. As we can see, below $H_c^{(1)}=D_+$, $\s_a=0$ and $\s_b\neq 0$,
while the situation is reversed above the spin-flop transition. The
component $\s_c$ is continuous at the spin-flop transition. In principle,
its slope as a function of the magnetic field changes at the spin-flop
transition according to Eqs.\ \pref{t<tn1} and \pref{t<tn2}. However, for
the parameter values used here, as appropriate for {\lco}, this change is
almost undistinguishable in Fig.\ \ref{phase-dia}. Moreover, also the
magnitude of the in-plane component is continuous at the transition, the
change being only in its direction.

As far as the second critical field is concerned, it turns out that Eq.\
\pref{hc2}, which uses the value of $\sigma_0$ at $H=0$, is an excellent
estimate of the second critical field $H_c^{(2)}$, obtained by means of the
self-consistent value $1-N\tilde I_\perp(H)$: indeed, since we found
$\s_0(T=0,H=0)=0.4$ (see inset), and $H_c^{(2,naive)}=77$ T, Eq.\
\pref{hc2} would give $H_c^{(2)}=30.8$ T, which is almost the value found
numerically, see Fig.\ \ref{phase-dia}. It is worth noting that $H_c^{(2)}$
has been measured recently in $1\%$ doped {\lasco} sample by Ono et
al. \cite{Ono}, who found $H_c^{(2)}\simeq 20$ T. Thus, even though
$H_c^{(2)}$ has not been measured in undoped {\lco}, where it is expected
to be larger than in the $1\%$ doped sample, the value of 30 T following
from Eq.\ \pref{hc2} seems more realistic than the bare estimate
\pref{hc2naive}, which gives 77 T. Observe that neglecting the quantum
renormalization of the order parameter in the estimate of the second
critical field can lead to an underestimate of the mass of the $c$ mode, as
it has been done in Ref.\ \onlinecite{Thio90}, where Eq.\ \pref{hc2naive}
has been used.

Fibally, we note that in the saddle-point approximation used so far the
transverse gaps are constant in temperature below $T_N$. However, one would
expect that a better approximation could reproduce the softening of the
transverse gaps as the temperature increases, as observed
experimentally. In this case also the value of the first critical field
$H_c^{(1)}$ would acquire a temperature dependence, which is instead absent
in the phase diagram of Fig.\ \ref{phase-dia}. Moreover, this could also
smoothen the discontinuity of $T_N$ at the spin-flop transition found at
saddle-point level.

\subsection{$\bH$ parallel to $c$}

\begin{figure}[htb]
\includegraphics[angle=0,scale=0.4]{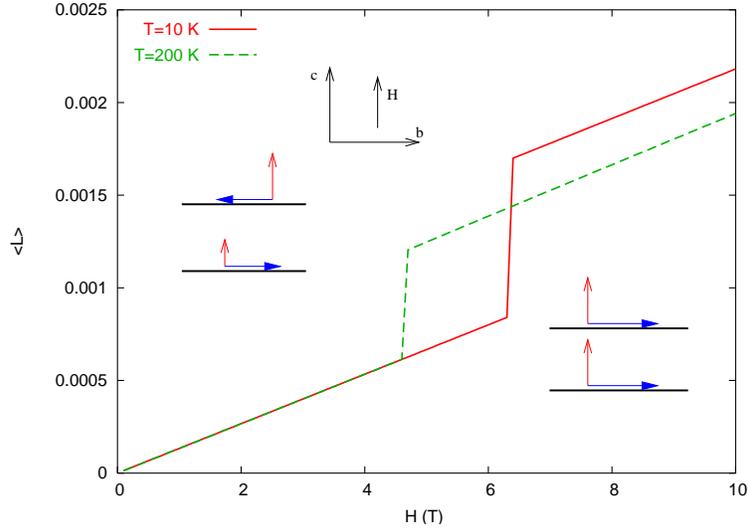}
\caption{(Color online) Magnetic-field dependence of the spin configuration
  and of the average uniform magnetization at two different temperatures
  for a field along the $c$ axis. The spin configuration is determined as
  usual by $\bS_i/S=(-1)^m e^{i\bQ \cdot {\bf r}_i}\bn^m+\bL^m$, where the
  staggered and uniform components are indicated by arrows with different
  tips as in Fig.\ \ref{fig-canting} and Fig.\ \ref{uniformb}.  At $H<H_c$,
  where $H_c=6$ T at $T=10$ K and $H_c=4.6$ at $T=200$ K, see also Fig.\
  \ref{fig-hcrit} below, the system has 3D antiferromagnetic order as
  described by Eq.\ \pref{sol1}. The difference between the $\bL_m$
  components (arrows with a two-lines tip) in neighboring planes is due to
  the DM-induced term $\bL_m^{DM}$ which changes sign from one layer to the
  next one, while $\bL_m^H$ is always parallel to ${\bf H}$, see Eq.\
  \pref{cunif}. Above the spin flop the spins are ordered ferromagnetically
  in neighboring layers, see Eq.\ \pref{sol2}, allowing for the uniform
  $\bL_m$ components to allign along the field in all the layers. The jump
  at the transition is proportional to the order parameter $\s_0$, and
  decreases as the temperature increases, see inset of Fig.\
  \ref{fig-hcrit}.}
\label{uniformc}
\end{figure}

When $\bH$ is along the $\hat\bx_c$ direction the last term in the action
\pref{nlsmh}  gives rise to an effective longitudinal staggered field:
\be
\lb{nlsmhpllc}
{\cal S}({\bf B})=
{\cal S}_0+\frac{1}{2gc}\sum_m\int \left[
  2iH(n_m^b\pd_\t n_m^a-n_m^a\pd_\t n_m^b)-H^2+H^2(n_m^c)^2]
+2h^mn_m^b+i\lambda_m 2gc(\bn_m^2-1) \right]. \ee
Following the line of the analysis performed in the previous section,
we first determine the ground-state configuration in the presence of the
magnetic field. Since also the effective staggered field is
longitudinal, we do not expect in this case to have a change of
direction of the equilibrium configuration. We can then look for a
ground-state solution of the form $\bn_m=\s_m^0\hat\bx_b$, which
gives:
\be
\lb{ssc}
\bar{\cal S}_{cl}=\frac{\b \cA}{2gc}\sum_m
\eta(\s_m^0-\s_{m+1}^0)^2+2h^m\s_m^0+m^2_m[(\s_m^0)^2-1].
\ee
The saddle-point equations then read
\bea
\eta(2\s_m^0-\s_{m+1}^0-\s_{m-1}^0)+h^m+m^2_m\s_m^0=0,\nn\\
\lb{sp} (\s_m^0)^2=1. \eea
As a consequence, two solutions are possible: (i) the order parameter is
the same in all the layers, and the ground-state action $\bar {\cal S}_{cl}$
vanishes:
\be
\lb{sol1}
\s_m^0=\s_0=- 1, \quad m^2=-(-1)^m h/\s_0=(-1)^m h, \quad \bar {\cal S}_{cl}=0.
\ee
This configuration is the same of the case without magnetic field. The
uniform spin components are:
\be
\lb{cunif}
\bL_m^{H}=\frac{H}{8aSJ}\hat\bx_c, \quad
\bL_m^{DM}=(-1)^m\frac{D_+}{8aSJ}|\s_0|\hat\bx_c,
\ee
so that the $\bL_m^{DM}$, which are ordered antiferromagnetically in
 neighboring layers (see Fig.\ \ref{fig-canting}), do not contribute to the
 average uniform magnetization $\langle \bL \rangle =\bL^H$ (see Fig.\
 \ref{uniformc}), but one takes advantage from the out-of-plane
 antiferromagnetic coupling; (ii) the order parameter change sign in
 neighboring layers, which means that the spins order {\em
 ferromagnetically} in the $c$ direction, with the moment $\bL_m^{DM}$
 oriented in the same direction:
\be \lb{cunif2} \bL_m^{H}=\frac{H}{8aSJ}\hat\bx_c, \quad
\bL_m^{DM}=\frac{D_+}{8aSJ}|\s_0|\hat\bx_c, \ee
giving $\langle \bL \rangle=(H+D_+|\s_0|)/8aSJ$.  This spin flop of the
uniform $c$ components of the spins leads to a lowering of the energy when
the gain in magnetic energy is larger than the cost coming from the
interlayer AF coupling. Observe that the average uniform magnetization
jumps discontinuosly at the spin-flop transition, the jump being
proportional to $\s_0$, so that it decreases as the temperature increases,
see Fig.\ \ref{uniformc}. 
The classical action in this configuration is:
\be
\lb{sol2}
\s_m^0=(-1)^m\s_0= -1, \quad m^2_m=m^2=-h/\s_0-4\eta=h-4\eta,
\quad \bar {\cal S}_{cl}=\frac{\b \cA N_l}{gc}\left(-h+2\eta\right),
 \ee
where $N_l$ is the number of layers. When $h>2\eta$ this second solution
becomes energetically favorable, so that the critical field for this spin-flop
transition is:
\be
\lb{hcrit}
H_c=\frac{2\eta }{D_+}.
\ee
When transverse fluctuations are included the first of Eqs.\
\pref{sp} will not change, while the value of the order parameter $\s_0$
will acquire quantum and thermal corrections due to transverse
fluctuations, according to:
\be
\lb{op}
\s_0^2=1-I_\perp(h/\s_0).
\ee
In Eq.\ \pref{op} we included the explicit dependence of the transverse
fluctuations on the order parameter $\s_0$ via the effective staggered
field $h$, as we will derive in detail below.
To determine $I_\perp$ we should distinguish the case below and above the
spin-flop transition. Let us first consider the case \pref{sol1} of low
field. Since the Lagrange multiplier is oscillating in neighboring layers,
it gives rise to a coupling between the transverse modes at $k_\perp$ and
$k_\perp+Q_\perp$, where $Q_\perp=\pi/d$. We then have:
$$
{\cal S}=\bar{\cal S}_{cl}+\frac{1}{2}\sum_{\a=a,c}\sum'_{\bk,k_\perp}
\Psi^+_\a(\bk,k_\perp)\hat{ G}_\a^{-1} \Psi_\a(\bk,k_\perp),
$$
where $\Psi^+_\a(\bk,k_\perp)=((n_\a(\bk,k_\perp),
n_\a(\bk,k_\perp+Q_\perp))$, the sum $\sum^{'}_{k_\perp}$ is over the reduced
Brilluoin zone for $k_\perp$, i.e $0\leq k_\perp\leq \pi/d$. The Green's
function for each mode is now a matrix. For the $a$ mode we find:
\be
\lb{gmaa}
\hat{ G}_a^{-1}=\frac{1}{gc}
\left( \begin{array}{cc}
\omega_n^2+\omega^2(\bk,k_\perp) &  m^2 \\
 m^2     & \omega_n^2+\omega^2(\bk,k_\perp+Q_\perp)
\end{array} \right),
\ee
with $\omega^2(\bk,k_\perp)=m_a^2+c^2\bk^2+2\eta(1-\cos k_\perp d)$ and
$m^2=-h/\s_0$, according to Eq.\ \pref{sol1}. The inverse Green's function
for the $c$ mode has an analogous expression, except that $m_a^2\ra
m_c^2+H^2$. After integration of the Gaussian fluctuations one obtains the
order-parameter equation \pref{op}, provided that Eq.\ \pref{gmaa} is used
to compute the fluctuations:
\be
\lb{mixprop}
\langle |n_a(\omega_n,\bk,k_\perp)|^2\rangle=
\frac{\omega_n^2+\omega^2(\bk, k_\perp+Q_\perp)}
{{\rm det} \, \hat G_a}.
\ee
The integral of the $a$-mode fluctuations has a structure similar to
the one of Eq.\ \pref{iahpllb}, with two contributions at the eigenvalues of
the matrix $\hat G_a$:
\be
\lb{iahpllc}
I_{a}=
\frac{1}{\b V}\sum_{\omega_n,\bk,k_\perp}
\langle |n_{a}(\omega_n,\bk,k_\perp)|^2\rangle=
\frac{1}{V}\sum_{\bk,k_\perp}
\frac{Z_+(k_\perp)}{2\omega_+(\bk,k_\perp)}\coth \frac{\b
  \omega_+(\bk,k_\perp)}{2} +
\frac{Z_-(k_\perp)}{2\omega_-(\bk,k_\perp)}\coth \frac{\b
  \omega_-(\bk,k_\perp)}{2},
\ee
where
\bea
\omega^2_{\pm}(\bk,k_\perp)&=&
\frac{\varepsilon_a^2(\bk,k_\perp)+\varepsilon_a^2(\bk,k_\perp+Q_\perp)}
{2}\pm
\sqrt{\left(\frac{\varepsilon_a^2(\bk,k_\perp)-\varepsilon_a^2(\bk,k_\perp+Q_\perp)}{2}\right)^2+
m^2 }=\nn\\
\lb{oapm}
&=&c^2\bk^2+m_a^2+2\eta \pm \sqrt{4\eta^2\cos^2 k_\perp d +m^2}.
\eea
Analogously to the case discussed in the previous section, the
spectral weights $Z_{\pm}$ of the two poles are not equivalent, and
determine the main character of the excitation. In this case we have:
\be \lb{zqpm} Z_{\pm}(k_\perp)=\mp
\frac{-\omega_{\pm}^2+\omega^2(\bk,k_\perp+Q_\perp)}
{\omega_+^2(\bk,k_\perp)-\omega_-^2(\bk,k_\perp)}= \mp\frac{2\eta\cos
k_\perp d \mp \sqrt{4\eta^2\cos^2 k_\perp d +m^2} } {2
\sqrt{4\eta^2\cos^2 k_\perp d +m^2}}. \ee
\begin{figure}[htb]
\includegraphics[angle=0,scale=0.4]{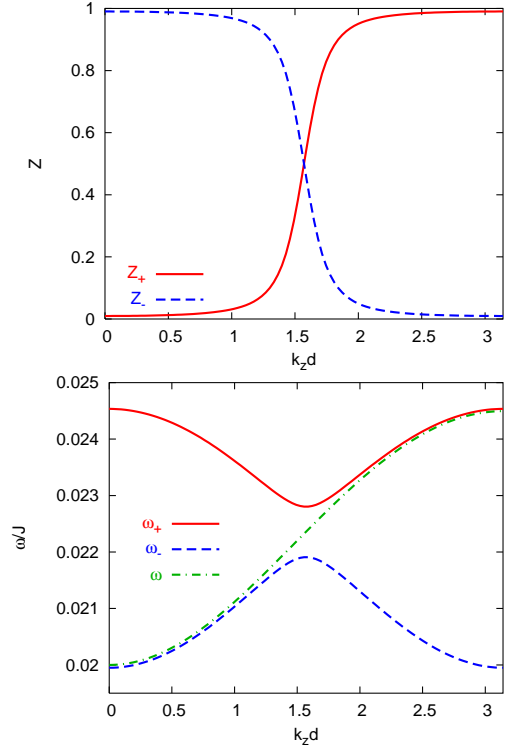}
\caption{(Color online) Low panel: out-of-plane momentum $k_z\equiv
  k_\perp$ dependence of the two solutions $\omega_\pm$ of Eq.\ \pref{oapm} at
  $\bk=0$, compared to the momentum dependence of the function
  $\omega(\bk=0,k_\perp)=\sqrt{m_a^2+2\eta(1-\cos k_\perp d)}$. We used
  $D_+=m_a=0.02 J$ and $H=1$ T$=10^{-3}$ J. Observe the small difference of
  $\omega_-$ with respect to $\omega$ at $k_\perp=0$. Top panel: momentum
  dependence of the spectral weight $Z_\pm(k_\perp)$ defined in Eq.\
  \pref{zqpm}. As one can see, at small momentum $Z_-\gg Z_+$, while at
  $k_\perp d> \pi/2$ the situation is reversed. As a consequence, the
  momentum sum in Eq.\ \pref{iahpllc} select always between $\omega_+$ and
  $\omega_-$ the solution which follows most closely the momentum dependence of
  the solution $\omega(k_\perp)$ of the case $H=0$.}
\label{spectral}
\end{figure}
The momentum dependence of $Z_\pm$ is reported in the top panel of Fig.\
\ref{spectral}, while in the lower panel the two solutions
$\omega_\pm(\bk=0,k_\perp)$ are plotted. As one can see, as $k_\perp$
increases the spectral weigh of the momentum sum in Eq.\ \pref{iahpllc}
moves from the solution $\omega_-(k_\perp)$ to the solution
$\omega_+(k_\perp)$, which follow closely the bare function
$\omega(k_\perp)$ in the two regimes $0<k_\perp d<\pi/2$ and $\pi/2<k_\perp
d<\pi$ respectively. The effect of the magnetic field is then twofold: it
affects the magnon gap $k_\perp=0$ and opens an additional one at $k_\perp
d=\pi/2$.  To compute explicitly the momentum sum in Eq.\ \pref{iahpllc} we
can observe that $Z_\pm(k_\perp)$ only depend on the out-of-plane momentum
$k_\perp$. We can then perform the usual integration over the in-plane
momentum $\bk$ in Eq.\ \pref{iahpllc}, obtaining an expression similar to
Eq.\ \pref{ia}-\pref{i3d}:
\be
\lb{ia3d}
I_a=
\frac{gT}{2\pi c}\int_{-\pi}^\pi \frac{dz}{2\pi}
Z_-(z)\ln\left\{
\frac{\sinh(c\Lambda/2T)}{\sinh A_-(z)/2T}\right\}+
Z_+(z)\ln\left\{
\frac{\sinh(c\Lambda/2T)}{\sinh A_+(z)/2T}\right\}, \quad H<H_c
\ee
where
\be
A_\pm(z)=\sqrt{m_a^2+2\eta\pm\sqrt{4\eta^2\cos z+(HD_+/\s_0)^2}}.
\ee
For the $c$ fluctuations one finds the same results, provided that $m_a^2\ra
m_c^2+H^2$ in all the above formulas.

Let us discuss now the issue of the magnon gaps. The spectral function of
the $a$ mode at $\bk=0,k_\perp=0$ has a two-poles structure
analogous to Eq.\ \pref{ca}, i.e.:
$$
 \cA_{a,c}(\omega>0)=
\left[\frac{Z_+(k_\perp=0)}{\omega_+}\d(\omega-\omega_+)+\frac{Z_-(k_\perp=0)}
{\omega_-}\d(\omega-\omega_-)\right]. 
$$
However, as observed before and shown in Fig.\ \ref{spectral}, at
$k_\perp=0$ is $Z_-/\omega_-\gg Z_+/\omega_+$, so that only the second term
contributes in the previous equation and allows us to identify the magnon
gap as the $\bk=0,k_\perp=0$ limit of the the function $\omega_-(\bk,k_\perp)$
above, i.e.:
\bea
\omega^2_a&=&m_a^2 +2\eta-\sqrt{4\eta^2+\left(\frac{ H
    D_+}{\s_0}\right)^2}, \nn\\
\lb{h<hc}
\omega^2_c&=&m_c^2 +H^2+2\eta-\sqrt{4\eta^2+\left(\frac{ H
    D_+}{\s_0}\right)^2}, \quad H<H_c,
\eea
which reduce to the conventional ones when $H=0$. Here we used explicitly
that $m^2=-h/\s_0$, according to Eq.\ \pref{sol1}.  Observe that this
result is quite different from the interpretation given in
Ref. \onlinecite{Papanicolaou}, where it was claimed that the acoustic and
optical modes are mixed. Instead, at $k_\perp=0$ only the mode $\omega_-$
is observed, as Raman measurements confirm.\cite{Gozar,theo-exp} Moreover, we
stress that according to the discussion below Eq.\ \pref{hplla}, for an
ordinary easy-axis AF we expected that $\omega_a$ is unchanged and
$\omega_c$ hardens for a field parallel to $c$. Instead, due to the
presence of the DM interaction, the two modes have a field dependence $\sim
-H^2 \gamma_{a,c}$, with $\gamma_a=D_+^2/(4\eta\s_0^2)$ and
$\gamma_c=-1+D_+^2/(4\eta \s_0^2)$. As a consequence, the $a$ mode always
softens, while the behavior of the $c$ mode depends on the ratio
$D_+^2/(4\eta\s_0)$.

\begin{figure}[htb]
\includegraphics[angle=-90,scale=0.4]{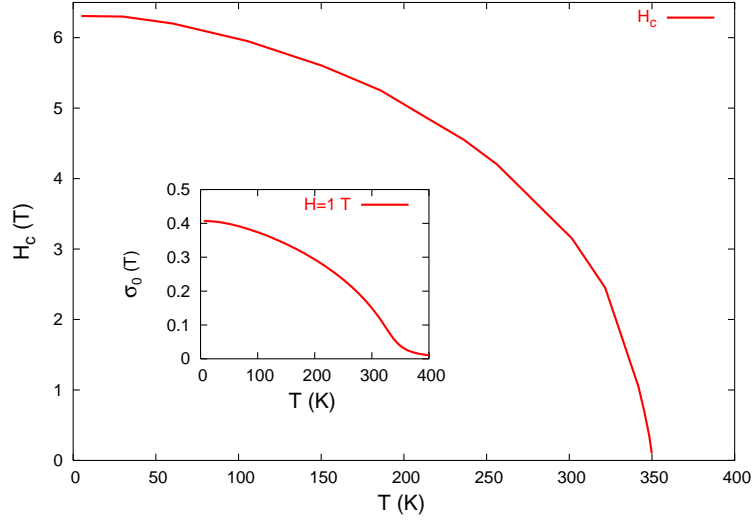}
\caption{(Color online) $H_c-T$ phase diagram for 
  $\bH \pll c$. We
  used the same parameter values as Fig.\ \ref{phase-dia}, except that in
  this case $g_s=2.4$, as appropriate for a field in the $c$ direction
  \cite{theo-exp}. The order parameter $\s_0(T)$ has been determined
  self-consistently trough Eq.\ \pref{op}, computing the transverse
  fluctuations with Eq.\ \pref{ia3d} and \pref{iaconv} below and above the
  spin-flop transition, respectively, where the critical field $H_c$ is
  defined trough Eq.\ \pref{hcritnew}. Inset: temperature dependence of the
  staggered order parameter at $H=1$ T.}
\label{fig-hcrit}
\end{figure}

Let us analyze now the case $H>H_c$. According to Eq.\ \pref{sol2} we find a
uniform saddle-point value of the constraint, $m^2_m=m^2=-h/\s_0-4\eta$, so
that the matrix of the transverse fluctuations is again diagonal in
momentum space:
\be
\lb{gm-hc}
\hat G^{-1}=
\left( \begin{array}{cc}
\omega_n^2+c^2\bk^2-2\eta(1+\cos k_\perp d)+m_a^2-h/\s_0 &  0 \\
 0       & \omega_n^2+c^2\bk^2-2\eta(1+\cos k_\perp d)+m_c^2+H^2-h/\s_0
\end{array} \right).
\ee
However, in the classical configuration \pref{sol2} the spins are order
ferromagnetically in neighboring planes, see Fig.\ \ref{uniformc}: this
means that the low-energy spin fluctuations are those at $k_\perp=\pi/d$ in
the notation of Eq.\ \pref{gm-hc}, so that the spin-wave gaps evolve at
$H>H_c$ in:
\bea
\omega^2_a&=&m_a^2 +\left( \frac{ H
    D_+}{|\s_0|}\right),\nn\\
\lb{h>hc}
\omega^2_c&=&m_c^2 +H^2+\left( \frac{ H
    D_+}{|\s_0|}\right), \quad H>H_c.
\eea
Since the matrix \pref{gm-hc} admits two simple poles, the transverse
fluctuations are described by the function \pref{i3d}, whith the masses
given by the previous equation:
\be
\lb{iaconv}
I_{a,c}=I_{3D}(\omega_{a,c}), \quad H>H_c
\ee
Observe that both for the case $H<H_c$ and $H>H_c$ the integral of
transverse fluctuations, given by Eqs. \pref{ia3d} and \pref{iaconv}
respectively, depend explicitly on the Lagrange multiplier $h/\s_0$, so
that Eq.\ \pref{op} is a self-consistency equation for the order parameter
at all the temperatures. This is quite different with respect to all the
cases analyzed before, where $I_\perp$ depends only on the transverse
masses and two separate regimes exist, $m^2=0, \s_0\neq 0$ below $T_N$, and
$m^2\neq 0$ and $\s_0=0$ above $T_N$ (see Eq.\ \pref{sce}). Here instead,
due to the effective longitudinal field $h^m$ in Eq.\ \pref{nlsmhpllc},
instead of the two regimes we obtain a single self-consistent equation
\pref{op} valid at all the temperatures. As a consequence, exactly as for a
ferromagnet in the presence of the external field, the order parameter
never vanishes, and the transition transforms into a crossover from a
regime where $\s_0$ is large to one where $\s_0$ is small \cite{MLVC}. This
is shown in the inset of Fig.\ \ref{fig-hcrit}, where we obtained $\s_0(T)$
at $H=1$ T by solving numerically the self-consistency equation \pref{op}
at various temperature. It is worth noting that once that transverse
fluctuations are included, also the definition of the critical field
\pref{hcrit} changes. Indeed, since in general $\s_0$ is lower than 1
(including $T=0$ due to quantum correction), the critical field becomes
itself a function of temperature. A first estimate of this effect can be
done by evaluating again the value of the action $\bar {\cal S}_{cl}$ in Eq.\
\pref{sol2} for a generic $\s_0$. We then obtain that the critical field
is:
\be
\lb{hcritnew}
H_c=\frac{4\eta}{D_+}\frac{\s_0}{(1+\s_0^2)}.
\ee
A more precise evaluation of $H_c$ could be done including also the
contribution of the Gaussian transverse fluctuations to the action
\pref{gmaa}. However, already Eq.\ \pref{hcritnew} allows one to recognize
that as the temperature increases the decrease of the order parameter
$\s_0$ induces also a decrease of the critical field for the spin-flop
transition. The resulting $H_c-T$ phase diagram, obtained by means of Eq.\
\pref{hcritnew} where $\sigma_0$ is the solution of the self-consistent
Eq.\ \pref{op}, is reported in Fig.\ \ref{fig-hcrit}.  

\begin{figure}[htb]
\includegraphics[angle=-90,scale=0.4]{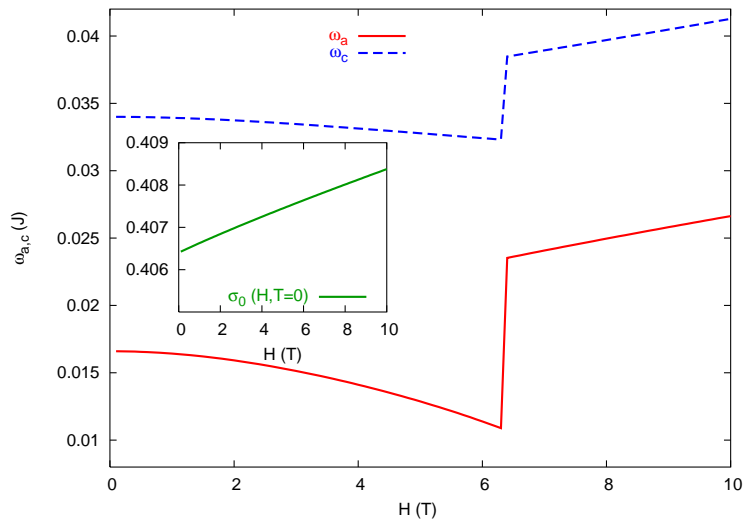}
\caption{(Color online) Field dependence of the magnon gaps at $T=0$ for
  $H\pll c$. Here $H_c\approx 6.5 $T, so below it the gap decrease
  according to Eq.\ \pref{h<hc}, while above it they increase according to
  Eq.\ \pref{h>hc}. The parameter values are the same of Fig.\
  \pref{fig-hcrit}. Inset: field dependence of the zero-temperature order
  parameter.}
\label{fig-gap-caxis}
\end{figure}

Once that we determined self-consistently the values of the order parameter
and of the critical field, we can also compute the field dependence of the
magnon gaps. In Fig.\ \pref{fig-gap-caxis} we show the field dependence at
$T=0$ of $\omega_a$ and $\omega_c$, as given by Eqs.\ \pref{h<hc} and
\pref{h>hc} below and above the spin-flop transition, respectively.  For
the interlayer coupling $\eta$ we used the value $\eta=8\times 10^{-5}
J^2$, which allows us to obtain a critical field at low temperature around
6.5 T, as the one measured experimentally \cite{Gozar}. Observe that this
value of $\eta$ is quite similar to the one obtained in Ref.\
\onlinecite{theo-exp} from the jump of the experimental measured in-plane gap at
$H_c$, even though such an estimate is done neglecting quantum corrections
to the order parameter. With this parameter values we find that the below
the spin-flop transition also the $c$ mode is slightly decreasing.

%
%

\subsection{$\bH$ parallel to $a$}
Finally, let us consider the case of a field along $\hat\bx_a$,
i.e. parallel to the direction of the DM vector $\bD_+$. In this case, the
last term of Eq.\ \pref{nlsmh} vanishes, and the system behaves as a
conventional easy-axis AF. As a consequence, Eq.\ \pref{hplla} holds,
giving a hardening of the $a$ gap and leaving the $c$ gap unchanged.

\section{Conclusions}

We have investigated the field dependence of the magnetic spectrum in
anisotropic two-dimensional and Dzyaloshinskii-Moriya layered
antiferromagnets. Starting from the appropriate spin-Hamiltonian for
each case, we obtained the magnon gaps and spectral intensities as a
function of the applied magnetic field, and we discussed the various
possible ground state configurations and phase diagram. In particular,
we showed that the peculiar coupling of the magnetic field with the
staggered order parameter induced by the DM interaction gives rise to
very interesting magnetic phenomena, such as spin-flop transitions and
rotation of spin quantization basis. The predictions of the theory
developed in this article are now ready to be compared with Raman 
spectroscopy experiments in {\srcuocl} and {\lco}, reported in the 
forthcoming article.\cite{theo-exp}

\section{Acknowledgements}

The authors would like to acknowledge invaluable discussions with
A. Lavrov.


\begin{thebibliography}{20}
\bibitem{coord-system}
We found it convenient here to use the $(abc)$ reference system,
rather than the $(bac)$ one used in a previous work.\cite{MLVC}
\end{thebibliography}
\end{document}